\documentclass[prd,twocolumn,showpacs,showkeys,preprintnumbers,floatfix,
nofootinbib,superscriptaddress]{revtex4-1}
\usepackage{amsfonts} 
\usepackage{amssymb} 
\usepackage{amsmath} 
\usepackage{graphicx} 
\usepackage{subfigure} 
\usepackage{array} 
\usepackage{dcolumn} 
\usepackage{bm} 
\usepackage{latexsym} 
\usepackage{longtable} 
\usepackage{hyperref} 
\graphicspath{{./figs/}}
\renewcommand{\tilde}{\widetilde}

\newcommand{\CA}{{\cal A}}

\newcommand{\CD}{{\cal D}}

\newcommand{\CL}{{\cal L}}
\newcommand{\CO}{{\cal O}}
\newcommand{\CS}{{\cal S}}

\newcommand{\MeV}{\mathop{\rm MeV}\nolimits}

\begin{document}


\title{Determining low-energy constants 
in partially quenched Wilson chiral perturbation theory}
\author{Maxwell T. Hansen}
\email[Email: ]{mth28@uw.edu}
\affiliation{
 Physics Department, University of Washington, 
 Seattle, WA 98195-1560, USA \\
}

\author{Stephen R. Sharpe}
\email[Email: ]{srsharpe@uw.edu}
%
%
\affiliation{
 Physics Department, University of Washington, 
 Seattle, WA 98195-1560, USA \\
}
%
%
%
%
%
%
\date{\today}
\begin{abstract}
In the low energy 
effective theory describing the partially quenched extension of
two light Wilson fermions, three low energy constants (LECs) appear in terms proportional to \(a^2\) (\(a\) being the lattice spacing). We propose methods to separately calculate these LECs, typically called \(W_6'\), \(W_7'\) and \(W_8'\).
While only one linear combination of these constants
enters into physical quantities, different combinations
enter into the description of the spectral density and
eigenvalue distributions of
the lattice Dirac operator and its Hermitian counterpart.
Thus it is useful to be able to determine the LECs
separately.
Our methods require studying certain correlation functions for either two or three pion scattering, which are accessible only in the partially quenched extension
of the theory.
\end{abstract}
\maketitle

\section{Introduction and Summary\label{sec:intr}}

Calculations using (improved versions) 
of Wilson fermions~\cite{Wilsonfermions}
have successfully approached~\cite{Luscherchiral}, 
and even reached~\cite{BMWphys}, 
physical light quark masses. The twisted-mass extension has
also been highly successful~\cite{TM4flavor}. 
Nevertheless, the explicit breaking of chiral symmetry
can lead to significant lattice artefacts that need to
be understood and controlled. For example, studying
the long-distance behavior of Wilson fermions using
chiral effective theory~\cite{SS,Creutz:1996bg}, 
one finds that, when quark
masses satisfy \(m\sim a^2\Lambda_{\rm QCD}^3\) 
(\(a\) being the lattice spacing),
discretization errors lead to a non-trivial phase diagram,
with one scenario (the ``first-order scenario'') having
a minimum pion mass, \(M_\pi^{\rm min} \sim a \Lambda_{\rm QCD}^2\),
and the other having a region of Aoki phase in which flavor
is spontaneously broken~\cite{Aokiphase}.
There have also been numerous studies of the properties of
mesons and baryons using the chiral effective theory---usually
called Wilson chiral perturbation theory (WChPT)---which provide
the functional forms needed to do simultaneous chiral and
continuum extrapolations.\footnote{%
For a recent review, see Ref.~\cite{goltermanleshouches}.}

In the unquenched theory with two light flavors
(a class of theories which includes physical QCD
 if one treats the strange quark as heavy), the chiral Lagrangian
 contains only one independent term proportional to \(a^2\). As a
 result, \(a^2\) corrections enter with a single low energy constant
 (LEC), denoted \(c_2\) in Ref.~\cite{SS}.
The sign of this constant determines the vacuum structure
when \(m\sim a^2\Lambda_{\rm QCD}^3\).
If one wants to go beyond the phase structure, however, and
use the chiral effective theory to determine discretization errors
in the spectral density
of the Hermitian Wilson-Dirac operator~\cite{gap06},
or, more generally, to determine the detailed properties
of low lying eigenvalues of the Wilson-Dirac operator~\cite{jac1,jac2},
then it turns out that one must consider the partially quenched (PQ)
extension of WChPT. In this extension, there are three LECs
entering in terms proportional to \(a^2\), denoted \(W_6'\), \(W_7'\) and \(W_8'\)
[see Eq.~(\ref{eq:lolag}) below], of which \(c_2\) is a particular linear combination [see Eq.~(\ref{eq:c2def})].
The detailed properties of the spectrum and eigenvalues
depend on all three LECs and not just on \(c_2\).
It is thus of interest to determine the
LECs separately, and this is the topic of the present work.

\begin{figure}[btp!]
{\includegraphics[angle=270,width=0.45\textwidth]{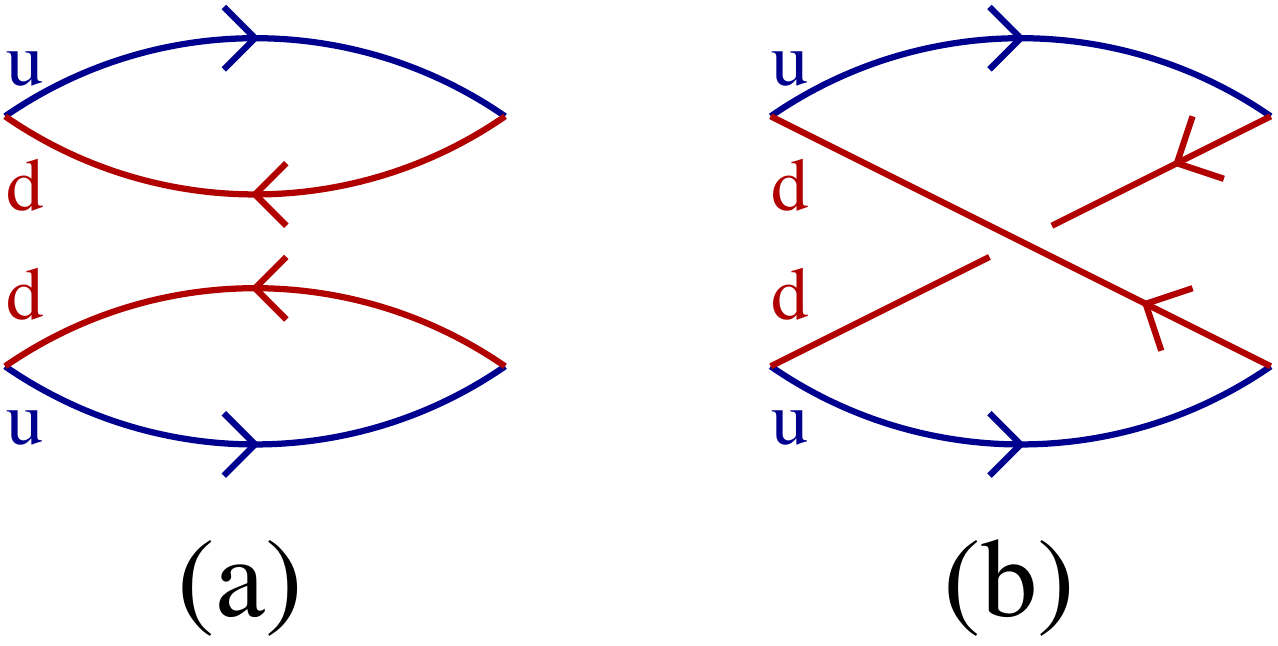}}
\caption{
Quark contractions contributing to \(\pi^+\pi^+\) scattering.
Contractions involving the interchange of 
the final state pions are not shown.
\label{fig:contr}}
\end{figure}

Our proposal builds upon one of the methods being
used to determine \(c_2\). This is to calculate certain pion scattering
lengths (i.e. the scattering amplitudes at threshold),
which in the continuum are proportional to \(m\),
but which also have
contributions proportional to \(c_2 a^2\) when discretization errors
are included~\cite{ABB}.
Computationally, the simplest choice is to
calculate the \(\pi^+\pi^+\to\pi^+\pi^+\) scattering amplitude,
for this involves no quark-antiquark annihilation contractions,
as illustrated in Fig.~\ref{fig:contr}.
The scattering length can be calculated 
from the energy shift \(\delta E=E(\pi^+\pi^+)-2 m_\pi\)
using  the method of Ref.~\cite{Luscher:1986pf}.
(More details of this method will be given below.)
Such a calculation is in fact presently being carried out~\cite{c2calc}.

Our method for calculating two of the LECs is to
consider separately the quark-disconnected contraction 
of Fig.~\ref{fig:contr}(a)
and the quark-connected contraction of Fig.~\ref{fig:contr}(b).
This separation is simple in a numerical simulation,
but it introduces a problem in the theoretical description.
To pick out the separate contractions requires
(for the two-flavor theory that we consider)
using a PQ theory. We stress that in
this application of partial quenching
the valence quarks are all degenerate with the sea quarks and
have the same action. This is in contrast with the most common
application in which the valence and sea quark masses
(and sometimes also actions) differ.

While well defined as a Euclidean field theory, the PQ theory
is unphysical, and the method of Ref.~\cite{Luscher:1986pf}
for determining the scattering lengths does not apply.
In particular, the individual contractions cannot be written as
a sum of exponentials, as is the case for their sum.
Our proposal is instead to {\em directly fit the 
correlation functions calculated in the simulation to
the predictions of PQWChPT}. 
These predictions depend, at leading order (LO),
on the LECs $W'_6$ and $W'_8$, and thus these two
constants can be determined from lattice data at 
sufficiently small $m$ and $a^2$. 
This method of comparison was first
introduced in the quenched theory~\cite{BGpipi}.
In both the quenched and PQ cases, the key point is that
the effective chiral theory reproduces, at long distances,
the unphysical nature of the underlying theory.

\begin{figure}[tbp!]
{\includegraphics[angle=270,width=0.48\textwidth]{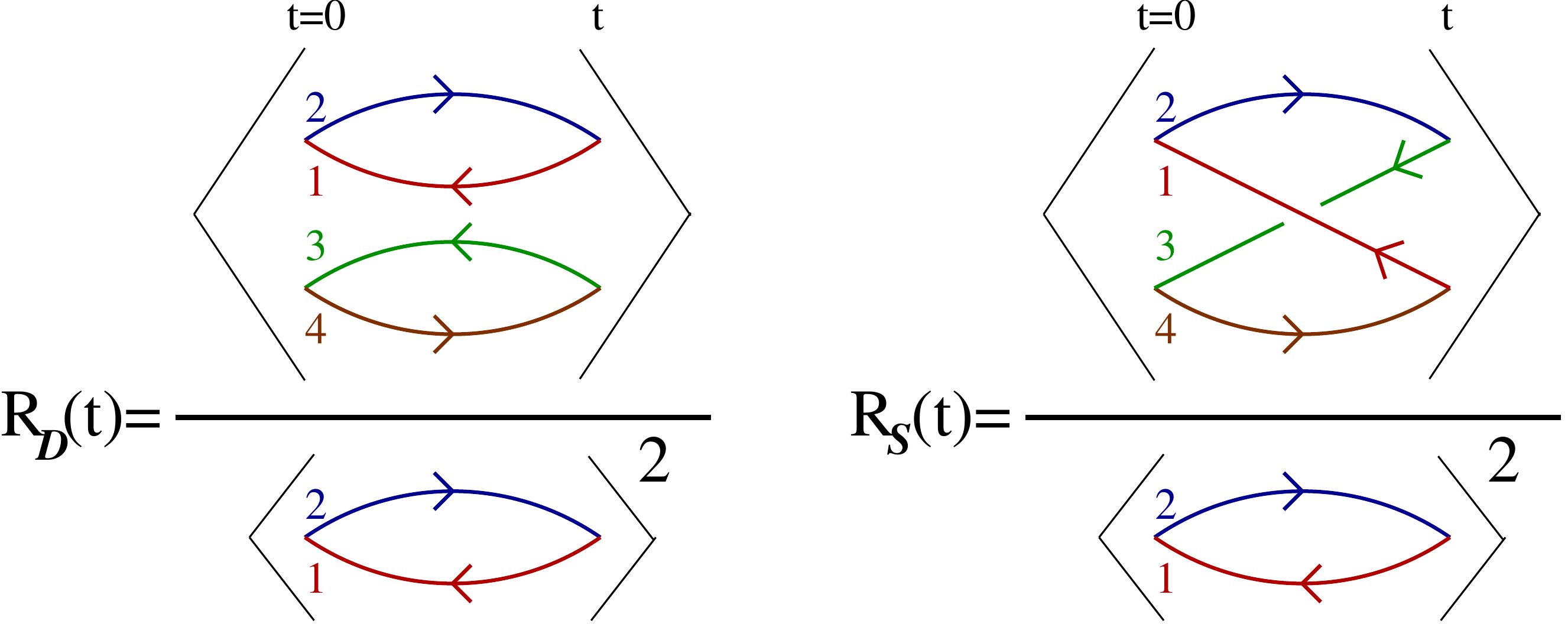}}
\caption{
Ratios of partially quenched correlation functions used to extract LECs.
Lines show quark propagators, with the valence flavor indicated by the
label [and color].
All interpolating operators are at zero three-momentum, and are placed at
the Euclidean times indicated.
Expectation values are taken with two sea quarks having the
same action and masses as the valence quarks.
\label{fig:ratios}}
\end{figure}

Specifically, we suggest calculating the ratios
of correlation functions shown in Fig.~\ref{fig:ratios}
[and defined in Eqs.~(\ref{eq:cd}-\ref{eq:Cpidef}) below].
This calculation must be done outside the Aoki phase, if present,
as we assume that flavor is not spontaneously broken.
As we show in Sec.~\ref{sec:finvol}, 
at LO in ChPT the ratios are given by
\begin{eqnarray}
R_{\CD}(t) &=& 1 + \mathcal O(1/L^3) + 2 w'_6\frac{t}{4 M_\pi^2 L^3} \,,
\label{eq:RDLO} \\
R_{\CS}(t) &=& \mathcal O (1/L^3) + \left(w'_8 - \frac{M_\pi^2}{f^2}\right) \frac{t}{4 M_\pi^2 L^3} \,,
\label{eq:RSLO}
\end{eqnarray}
where $w'_{6,8}$ are proportional to $W'_{6,8}$ [see Eq.~(\ref{eq:lilwdef})],
and $f\approx f_\pi$.
Thus these two LECs can be determined from the terms linear in $t$.
This requires that $t$ be large enough that contributions from excited
pion states, which fall exponentially, can be neglected.
It also requires that the linear terms in \(t\) can be distinguished from quadratic terms which appear at higher order in ChPT and which scale as \(1/L^6\).
In practice, these conditions seem reasonable 
(see, e.g., Ref.~\cite{c2calc}).

In a ratio of physical correlation functions, the 
contribution linear in \(t\) is
the first term in the expansion of the exponential $\exp(-\delta E t)$.
In the PQ theory, by contrast, neither $R_{\CD}(t)$ nor $R_{\CS}(t)$
are exponentials. This is immediately clear for $R_{\CS}(t)$ due to
the absence of an \(L\)-independent constant term and will be shown explicitly
for $R_{\CD}(t)$ by calculating 
the quadratic term---see Eq.~(\ref{eq:RDfinal}).
Nevertheless, as long as it is possible to pick out the term linear 
in \(t\) one can extract the LECs.

In practice, subleading terms in the chiral expansion
are significant at the values of $m$ and $a^2$ used in present simulations.
Thus we have extended the calculation of the ratios
to next-to-next-to-leading order (NNLO)
in the power counting appropriate to the \(m\sim a^2\) regime
(usually called the ``large cut-off effects'' or LCE regime). 
This power-counting is explained in Sec.~\ref{sec:WChPT}.
It differs from the usual continuum
power-counting in that one-loop effects are of NNLO, rather than 
next-to-leading order (NLO).
The only NLO contributions are from analytic terms.

The LO results are generalized in a fairly simple way. To describe
this, we first define \(\CD(s,t,u)\) and \(\CS(s,t,u)\) as the
infinite volume, PQ scattering amplitudes corresponding to the
contractions of Figs.~\ref{fig:contr}(a) and (b) respectively [see
Eqs.~(\ref{eq:Ddef}) and (\ref{eq:Sdef}) below]. We then observe that
the coefficients of \(t/4 M_\pi^2 L^3\) in Eqs.~(\ref{eq:RDLO}) and
(\ref{eq:RSLO}) are simply the LO values of \(\CD(4M_\pi^2,0,0)\) and
\(\CS(4M_\pi^2,0,0)\), the amplitudes at threshold.
What we show in Sec.~\ref{sec:finvol} is that the
coefficients of $t/(4M_\pi^2 L^3)$, when evaluated to NNLO 
in the chiral expansion, continue to equal the infinite-volume 
threshold amplitudes, also evaluated to that order.\footnote{%
This is true up to
finite-volume corrections proportional to $\exp(-M_\pi L)$,
which are generically present also in unquenched applications,
and are usually small in actual simulations.}
In effect, picking out the coefficient of the term linear in $t$ in the
ratios is, for the threshold amplitude, like performing LSZ reduction. 

The full NNLO
results are given in Eqs.~(\ref{eq:RDfinal}) and (\ref{eq:RSfinal}).
We display here only simplified forms which show the essential features:
\begin{align}
\begin{split}
R_{\CD}(t) &= 1 + \CO\left(\frac{M_\pi^2}{f^2}\frac{1}{M_\pi^3 L^3}\right) \\
&{} + \CD(4 M_\pi^2,0,0) \frac{t}{4 M_\pi^2 L^3} \left[1 + \CO\left(\frac{M_\pi^2}{f^2}\frac{1}{M_\pi L}\right) \right] \\
&{}  \hspace{-30pt} + {\cal O}\left(\left[\frac{M_\pi^2}{f^2} \frac{t}{M_\pi^2 L^3}\right]^2\right) + {\cal O}(e^{-M_\pi L}) + \mathrm{exp. suppr.} \,,
\end{split} \label{eq:RDNNLO} \\
\notag \\
\begin{split}
R_{\CS}(t) &= \CO\left(\frac{M_\pi^2}{f^2}\frac{1}{M_\pi^3 L^3}\right) \\
&{} + \CS(4 M_\pi^2,0,0) \frac{t}{4 M_\pi^2 L^3} \left[1 + \CO\left(\frac{M_\pi^2}{f^2}\frac{1}{M_\pi L}\right) \right] \\
&{} \hspace{-30pt}+ {\cal O}\left(\left[\frac{M_\pi^2}{f^2} \frac{t}{M_\pi^2 L^3}\right]^2\right) + {\cal O}(e^{-M_\pi L}) + \textrm{exp. suppr.} \,,
\end{split} \label{eq:RSNNLO}
\end{align}
where \(\CD\) and \(\CS\) are evaluated to NNLO, and the results at
threshold are given in Eqs.~(\ref{eq:CDthresh}) and
(\ref{eq:CSthresh}). In our ``big \(\CO\)'' notation, \(M_\pi^2/f^2\)
may also stand for any of the terms proportional to \(a^2 W'_k\), as
these are of the same order in the LCE regime. 
By ``exp. suppr.'' we mean contributions which fall off
exponentially with $t$, e.g. due to excited states.

As can be seen from Eqs.~(\ref{eq:RDNNLO}) and (\ref{eq:RSNNLO}) there
are three expansions being used. First, there is the usual chiral
expansion (supplemented by powers of $a$)
in the expressions for $\CD$ and $\CS$.
Second, there is an expansion in powers of \(t\), as indicated in the
last lines of (\ref{eq:RDNNLO}) and (\ref{eq:RSNNLO}). Finally, for
each power of \(t\) there is a sequence of subleading terms, as is
shown in the middle line of each equation.

Our calculation in Sec.~\ref{sec:finvol} yields
the first non-trivial correction in each of these expansions,
namely the chiral corrections to $\CD$ and $\CS$ and the
contributions to the ratios proportional to $t^2/L^6$ and $t/L^4$.
The latter two  are given explicitly in
(\ref{eq:RDfinal}) and (\ref{eq:RSfinal}).
We stress again that, if the numerators in the ratios
were physical correlation functions,
then, using the results of Ref.~\cite{Luscher:1986pf}, one would find
that the $t/L^4$ and $t^2/L^6$ terms would be proportional to the
square of the threshold scattering amplitude.
This is not true in the PQ theory, but 
one can nevertheless calculate these terms.

With the NNLO results (\ref{eq:RDNNLO}) and (\ref{eq:RSNNLO}) in hand,
we can discuss our proposal for determining LECs in more detail.
In order for the term linear in $t$ to dominate
over the quadratic term, it must satisfy
\begin{equation}
t \ll {f^2 L^3} \sim \frac{(M_\pi L)(f L)^2}{M_\pi} \,.
\label{eq:noquadcond}
\end{equation}
Since $M_\pi L \gg 1$ and $f L \gtrsim 1$  to avoid large finite-size
effects, one sees that the constraint on $t$ is rather weak. 
One may also try and fit the ratios including
the quadratic term, and in this regard we note that the 
coefficient of $t^2$
is given by a linear combination of $M_\pi^2/f^2$ and 
the LECs $W_6'$ and $W_8'$, so that no new parameters are needed.

Assuming that one can determine the coefficient of $t$, one must
disentangle its dependence on $M_\pi^2$ and on $1/L$ in order
to extract the LECs $W'_6$ and $W'_8$. Here it is helpful
that the $1/L$ correction itself depends on these same two LECs.
At the least, this will allow an {\em a posteriori}
estimate of the size of the $1/L$ correction.

Finally, one must do a chiral extrapolation of the resulting
threshold amplitudes, attempting to pick out the LO terms which
give the desired LECs. From 
Eqs.~(\ref{eq:CDthresh}) and (\ref{eq:CSthresh}), the
general chiral behavior is, schematically,
\begin{eqnarray}
\CD,\CS &\sim& a^2 + M_\pi^2 + a^3 + a M_\pi^2 + a^4 (1+ \log M_\pi ) 
\nonumber\\
&& + a^2 M_\pi^2 (1 + \log M_\pi)  + M_\pi^4 (1 + \log M_\pi)
\,.
\end{eqnarray}
The coefficients of the chiral logarithms are either fixed
(for the continuum logarithm) or given in terms of all
three $W'_k$ (for the logarithms multiplying factors of $a$).
The analytic terms involve many other LECs, however, both from the
continuum theory and induced by discretization errors.
Thus it seems very unlikely that one will be able do more
than a fit to the generic form given above and extract the
LO $a^2$ and $M_\pi^2$ terms.
This would allow a determination of
$W'_6$ and $W'_8$, but not $W'_7$.

In light of this, we have devised an alternative method for
determining \(W_7'\), in which it contributes at tree-level.
This requires studying a particular three-pion correlation function.
Since this  will be challenging to implement in a simulation,
we describe the method only briefly in an appendix.

We close this section by noting that other methods 
for determining $W'_6$, $W'_7$ and $W'_8$ have recently been proposed.
These use the eigenvalue distributions of the Hermitian
Wilson-Dirac operator with Wilson-like fermions
(which are sensitive to all three LECs)~\cite{jac1,jac2}, 
the masses of pions and the scalar correlator
in a mixed-action simulation with
overlap valence quarks and twisted-mass sea quarks
(which can determine $W'_6$ and $W'_8$)~\cite{Herdoiza11},
and the mass of the quark-connected neutral pion with twisted-mass
quarks (which determines $W'_8$)~\cite{HS}.\footnote{%
It has also been shown in Refs.~\cite{jac1,jac2,HS} 
that $W'_8$ is necessarily negative.}
We think that pinning down the LECs will not be easy, and hope the
method proposed here can contribute along with these other approaches.

The remainder of this article is organized as follows.
In the following section, we give a brief recapitulation of the
pertinent details of PQWChPT, including
the power-counting of the LCE regime.
In Sec.~\ref{sec:infvol} we present our results for
the infinite-volume PQ scattering amplitudes, which we do for
general momentum, and at NNLO in the LCE power counting.
The description and calculation of the
finite-volume correlations functions are presented
in Sec.~\ref{sec:finvol}, 
which forms the core of the technical part of this paper.
We include two appendices.
The first contains details concerning analytic NLO and NNLO contributions
to the scattering amplitudes.
The second describes our proposal for determining $W'_7$.

\section{Partially Quenched Wilson Chiral Perturbation Theory}
\label{sec:WChPT}

In this section we define the required PQ
scattering amplitudes and recall the essentials of WChPT
for the partially quenched theory.
We consider a theory with two sea quarks, and introduce four
valence quarks, and their corresponding ghosts, in order to
define the desired amplitudes. All quarks and ghosts are 
degenerate with mass \(m\). With this quark content, the chiral field
\begin{equation}
\Sigma = \exp \left( \sqrt{2} i \pi / f\right)
\end{equation}
lies in the graded group \(SU(6\vert4)\).
We work in the LCE regime where \(m\sim a^2 \), so
the Lagrangian is broken into leading and subleading parts as 
follows~\cite{ABB}:
\begin{equation}
\label{eq:powcou}
\begin{split}
{\mathrm{LO}} &:\ p^2, m, a^2
\\
{\mathrm{NLO}} &:\ p^2 a, ma, a^3
\\
{\mathrm{NNLO}} &:\ p^4, p^2 m, m^2, p^2 a^2, ma^2, a^4  \,.
\end{split}
\end{equation}

The partially quenched scattering amplitudes of interest
correspond to the two contractions shown in Fig.~\ref{fig:contr}.
We label these, respectively, as $\CD$ and $\CS$ for double and single,
referring to the number of loops appearing in quark-flow diagrams.
The four valence flavors that we have introduced 
allow us to separate the contractions as illustrated 
in Fig.~\ref{fig:ratios}. The precise definitions are
\begin{multline}
\mathcal{D}(s,t,u) (2 \pi)^4 \delta^4(p+k-p'-k') = \\
\langle \pi_{12}(p') \pi_{21}(-p) \pi_{34}(k') \pi_{43}(-k)\rangle_{conn,amp}
\label{eq:Ddef}
\end{multline}
and
\begin{multline}
\mathcal{S}(s,t,u) (2 \pi)^4 \delta^4(p+k-p'-k') = \\
\langle \pi_{12}(p') \pi_{23}(-p) \pi_{34}(k') \pi_{41}(-k)\rangle_{conn,amp}
\,,
\label{eq:Sdef}
\end{multline}
where \(conn\) indicates that only pion-connected contributions are included and \(amp\) indicates standard amputation of external
propagators.
The subscripts on the pion field indicate their valence flavor and
$s$, $t$ and $u$ are standard Mandelstam variables,
\begin{align}
s=-(p+k)^2, \hspace{10pt} t=-(p-p')^2, \hspace{10pt} u=-(p-k')^2  \,,
\end{align}
with $p$, $k$, $p'$ and $k'$ all Euclidean four-vectors. 

Observe that the term ``amplitude'' only applies loosely here because
of the unphysical partial quenching. \(\mathcal{D}\) and
\(\mathcal{S}\) do not satisfy unitarity constraints and only certain
linear combinations give physical amplitudes.
In particular, the relation of PQ amplitudes to the amplitude for
\(\pi^+\) scattering 
is
\begin{multline}
\label{eq:apiplus}
\mathcal{A}_{\pi^+}(s,t,u) = \\ \mathcal{D}(s,t,u) +
\mathcal{D}(s,u,t) + \mathcal{S}(s,t,u) + \mathcal{S}(s,u,t) \,.
\end{multline}
This is found by comparing quark-level Wick contractions. We use this
result to provide a partial check of our results for
\(\mathcal{D}\) and \(\mathcal{S}\) by doing an independent computation of
\(\mathcal{A}_{\pi^+}\) in \(SU(2)\) WChPT.

We can also relate the PQ amplitudes to the general physical
scattering amplitude. We recall that, for
\(\pi_i + \pi_k \rightarrow \pi_l + \pi_m\), 
with \(i,k,l,m \in \{1,2,3\}\), one can write
\begin{multline}
\label{eq:apireal}
\mathcal{A}_{ik \rightarrow lm}(s,t,u) = \\ \delta_{ik} \delta_{lm}
\mathcal{A}(s,t,u) + \delta_{il} \delta_{km} \mathcal{A}(t,s,u) +
\delta_{im} \delta_{lk} \mathcal{A}(u,t,s) \,.
\end{multline}
The amplitude $\CA$ is related to our amplitudes by
\begin{multline}
\label{eq:adef}
\mathcal{A}(s,t,u) = \\ \mathcal{D}(t,s,u) - \mathcal{S}(s,t,u) +
\mathcal{S}(u,t,s) + \mathcal{S}(t,s,u) \,.
\end{multline}
The result for \(\mathcal{A}\) at NNLO in WChPT is obtained
in Ref.~\cite{ABB}, and this relation
allows us to compare our results to those in that work.

In order to calculate \(\mathcal{D}\) and \(\mathcal{S}\) to
NNLO we must include all tree level and one
loop diagrams generated by the LO Lagrangian (\(\mathcal{L}_{LO}\))
but only the tree level diagrams from \(\mathcal{L}_{NLO}\) and
\(\mathcal{L}_{NNLO}\). 
The LO Lagrangian is~\cite{SS,BRS03}
\begin{equation}
\label{eq:lolag}
\begin{split}
\mathcal{L}_{\rm LO} & = \frac{f^2}{4} \langle\partial_\mu \Sigma
\partial_\mu \Sigma^\dagger\rangle - \chi\frac{f^2}{4} \langle \Sigma
+ \Sigma^\dagger \rangle \\ & {} - \hat{a}^2 W_6' \langle \Sigma +
\Sigma^\dagger\rangle^2 \\ & {} - \hat{a}^2 W_7'\langle\Sigma -
\Sigma^\dagger\rangle^2 \\ & {} - \hat{a}^2 W_8' \langle \Sigma^2 +
(\Sigma^\dagger)^2 \rangle \,.
\end{split}
\end{equation}
Here angle brackets indicate a supertrace (strace), and
\(\chi = 2 B_0 m\) and \(\hat a = 2 W_0 a\), with \(B_0\) and \(W_0\) 
leading-order LECs.
The term linear in \(a\) has been removed by the standard
redefinition of \(m\)~\cite{SS}. 
We use the ``small \(f\)'' convention in which \(f_\pi \approx 93\MeV\).

When the field \(\Sigma\) is restricted to \(SU(2)\), the number of
independent terms in \(\mathcal{L}_{\rm LO}\) is reduced. In particular,
the \(W_7'\) term vanishes and the \(W_6'\) and \(W_8'\) terms become
proportional. As a result, the unquenched amplitudes defined in
(\ref{eq:apiplus})-(\ref{eq:adef}) can only depend on \(W_6'\) and
\(W_8'\) through the combination \(2 W_6' + W_8'\). A convenient
definition for this, used in Ref.~\cite{ABB}, is
\begin{equation}
\label{eq:c2def}
c_2 = -32(2W_6' + W_8')\frac{W_0^2}{f^2}  \,.
\end{equation}

The NLO and NNLO Lagrangians are~\cite{GL84,BRS03}
\begin{align}
\begin{split}
\mathcal{L}_{\rm NLO} & = \hat a W_4 \langle \partial_\mu \Sigma \partial_\mu \Sigma^\dagger \rangle \langle \Sigma + \Sigma^\dagger\rangle \\
& {} + \hat aW_5 \langle \partial_\mu \Sigma \partial_\mu \Sigma^\dagger (\Sigma + \Sigma^\dagger) \rangle \\
& {} - \hat a \chi W_6 \langle \Sigma + \Sigma^\dagger \rangle^2 \\
& {} - \hat a \chi W_7 \langle \Sigma - \Sigma^\dagger \rangle^2 \\
& {} - \hat a \chi W_8 \langle \Sigma^2 + (\Sigma^\dagger)^2 \rangle \\
& {} + a^3 \mathrm{-terms}
\end{split} \\ \notag \\
\begin{split}
\mathcal{L}_{\rm NNLO} & = -L_1 \langle \partial_\mu \Sigma \partial_\mu \Sigma^\dagger \rangle^2 \\
& {} - L_2 \langle \partial_\mu \Sigma \partial_\nu \Sigma^\dagger \rangle \langle \partial_\mu \Sigma \partial_\nu \Sigma^\dagger \rangle \\
& {} - L_3 \langle \partial_\mu \Sigma \partial_\mu \Sigma^\dagger \partial_\nu \Sigma \partial_\nu \Sigma^\dagger \rangle \\
& {} + \chi L_4 \langle \partial_\mu \Sigma \partial_\mu \Sigma^\dagger \rangle \langle \Sigma + \Sigma^\dagger\rangle \\
& {} + \chi L_5 \langle \partial_\mu \Sigma \partial_\mu \Sigma^\dagger (\Sigma + \Sigma^\dagger) \rangle \\
& {} - \chi^2 L_6 \langle \Sigma + \Sigma^\dagger \rangle^2 \\
& {} - \chi^2 L_7 \langle \Sigma - \Sigma^\dagger \rangle^2 \\
& {} - \chi^2 L_8 \langle \Sigma^2 + (\Sigma^\dagger)^2 \rangle \\
& {} - L_{PQ} \mathcal{O}_{PQ} \\
& {} + a^2 m \mathrm{-terms} \\
& {} + a^2 p^2 \mathrm{-terms} \\
& {} + a^4 \mathrm{-terms}  \,,
\end{split}
\end{align}
where~\cite{OPQ}
\begin{equation}
\begin{split}
\mathcal{O}_{PQ} & = \langle \partial_\mu \Sigma \partial_\nu \Sigma^\dagger \partial_\mu \Sigma \partial_\nu \Sigma^\dagger \rangle \\
& {} + 2 \langle \partial_\mu \Sigma \partial_\mu \Sigma^\dagger \partial_\nu \Sigma \partial_\nu \Sigma^\dagger \rangle \\
& {} - (1/2)\langle \partial_\mu \Sigma \partial_\mu \Sigma^\dagger \rangle^2 \\
& {} - \langle \partial_\mu \Sigma \partial_\nu \Sigma^\dagger \rangle \langle \partial_\mu \Sigma \partial_\nu \Sigma^\dagger \rangle  \,.
\end{split}
\end{equation}
The $a^3$, $a^2 m$, $a^2 p^2$ and $a^4$ terms are discussed in
appendix~\ref{app:indcon}.

As above, restriction to \(\Sigma \in SU(2)\) reduces the number of
terms in the Lagrangian. As a result, physical amplitudes can only
depend on the following combinations of NLO and NNLO LECs,
\begin{gather}
\label{eq:phys68}
2W_6' + W_8',\ \ 2 W_6 + W_8,\ \ 2 L_6 + L_8, \\
2 W_4 + W_5,\ \ 2 L_4 + L_5,\\
2 L_1 + L_3,\ \ L_2, 
\end{gather}
as well as on the LECs from \(a^3\), \(a^2 m\), \(a^2 p^2\) and \(a^4\) terms.
In particular, physical amplitudes
must be independent of \(W_7'\), \(W_7\), \(L_7\) and \(L_{PQ}\) because
the associated Lagrangian terms vanish when there are only two
quarks.

\section{Infinite Volume Partially Quenched Scattering Amplitudes}
\label{sec:infvol}

\begin{figure}[tbp!]
{\includegraphics[angle=270,width=0.45\textwidth]{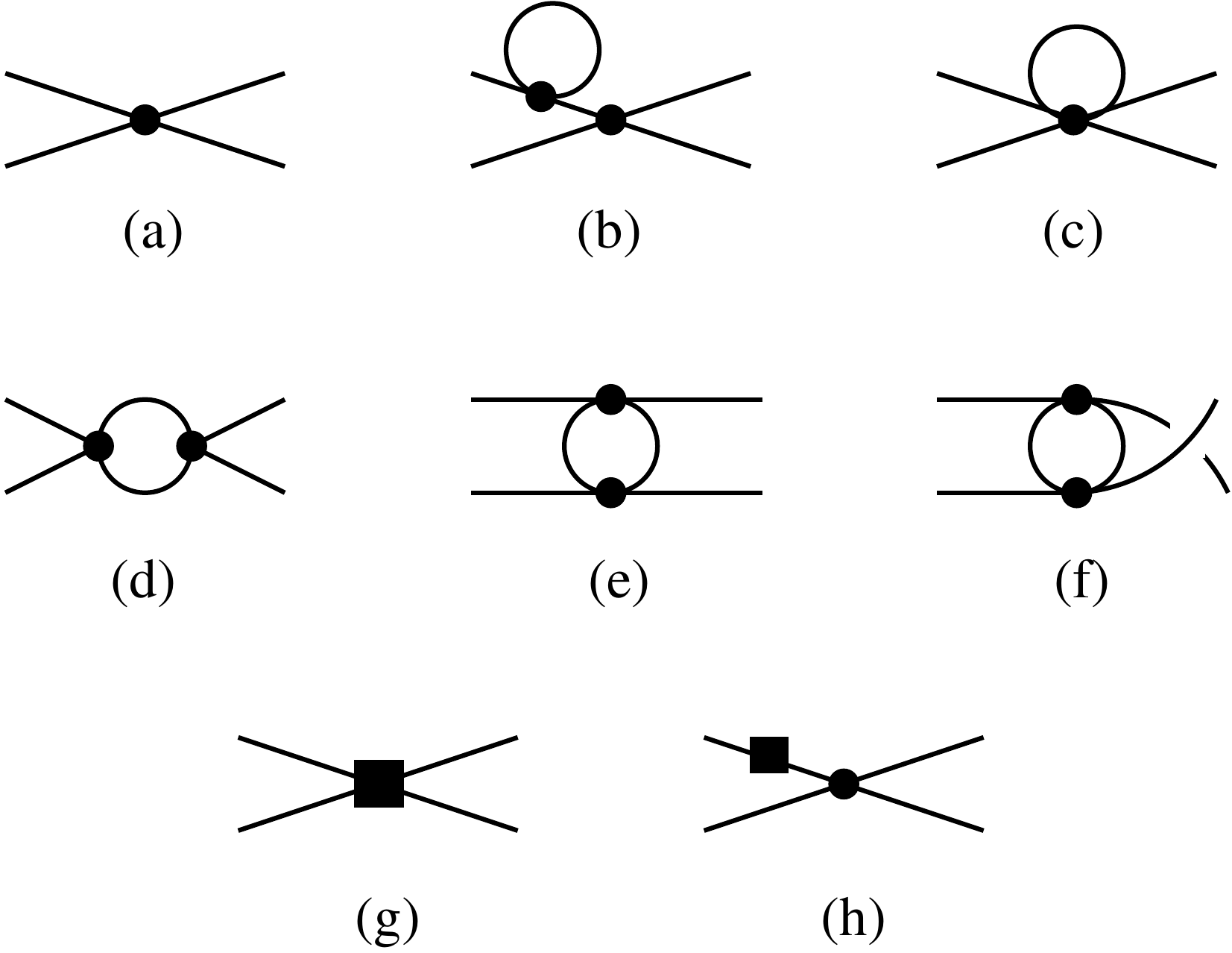}}
\caption{
Classes of diagrams contributing to the PQ scattering amplitudes through NNLO.
Filled circles represent vertices from $\CL_{\rm LO}$ while
filled squares represent vertices from $\CL_{\rm NLO}$ and
$\CL_{\rm NNLO}$.}
\label{fig:infvol}
\end{figure}

In this section we calculate the on-shell PQ amplitudes through NNLO.
Although, as noted earlier, the PQ theory is defined in Euclidean space,
we can nevertheless analytically continue the amplitudes to
Minkowski momenta and set them on shell. It turns out that it
is these on-shell amplitudes which appear in the coefficient
of $t$ in the ratios of Fig.~\ref{fig:ratios}.

The diagrams contributing through NNLO are
shown in Fig.~\ref{fig:infvol}. At LO, only Fig.~\ref{fig:infvol}(a)
contributes, and leads to the following results for the on-shell amplitudes:
\begin{align}
\label{eq:dtree}
\mathcal{D}^{(0)} & = 2 w_6' \\
\label{eq:stree}
\mathcal{S}^{(0)}(s) & = w_8' 
+ \frac{1}{2f^2} \left(2 M_0^2 - s\right) .
\end{align}
Here, $M_0$ is the LO pion mass, given by
\begin{equation}
M_0^2 = \chi + f^2(2w_6' + w_8') \,,
\end{equation}
where we have introduced rescaled, dimensionless LECs
\begin{equation}
\label{eq:lilwdef}
w_k' = \frac{16 \hat a^2 W_{k}'}{f^4} \hspace{10pt} k=6,7,8.
\end{equation}
Note that, because all quarks and ghosts are degenerate, 
all pions have the same mass, whatever their composition.

Combining Eqs.~(\ref{eq:dtree}) and (\ref{eq:stree}) according to
(\ref{eq:adef}) and using the on-shell result
\(s+t+u = 4M_0^2\), we find
\begin{equation}
\label{eq:atree}
\mathcal{A}^{(0)}(s) = \frac{s-M_0^2}{f^2} + 2 w_6' + w_8' \,.
\end{equation}
This agrees with the result of Ref.~\cite{ABB}.

We now turn to higher-order contributions,
considering first the loop graphs of NNLO
and then turning to the analytic contributions of NLO and NNLO.
Non-analytic contributions to
mass and wavefunction renormalization arise from the ``tadpole''
diagrams exemplified by Fig.~\ref{fig:infvol}(b).
Including the results from these diagrams
(and also the analytic analogs in Fig.~\ref{fig:infvol}(h)), 
the form of the valence-valence propagator near the physical pole is
\begin{equation}
\langle \pi_{ij}(p) \pi_{ji}(-p) \rangle \sim \frac{1}{(1 + \delta
z)(p^2 + M_\pi^2)} \,,
\end{equation}
with \(\sim\) indicating that the two sides differ by terms 
regular at the pole. This expression defines both the wave function
renormalization \(\delta z\) and the physical mass \(M_\pi\). It is
further convenient to define \(M_\pi^2 = M_0^2 + \delta M^2\).

Evaluating the tadpole diagrams we find
\begin{align}
\label{eq:deltmloop}
\delta M_{loop}^2 & = \left[\frac{1}{2} M_0^2 -
\frac{5}{2}f^2(2w_6'+w_8') \right]I_1 \\
\label{eq:deltzloop}
\delta z_{loop} & = -\frac{2}{3} I_1 \,,
\end{align}
where
\begin{equation}
I_1 = \frac{1}{f^2} \int_q \frac{1}{q^2 + M_0^2} \rightarrow
\frac{M_0^2}{16 \pi^2 f^2} \log \left(M_0^2/\mu^2 \right)
\end{equation}
is the standard tadpole integral. Here
\begin{equation}
\int_q = \mu^{\varepsilon} \int \frac{d^{4-\varepsilon} q}{(2
\pi)^{4-\varepsilon}}
\end{equation} 
and the arrow indicates evaluation in a modified minimal subtraction
scheme (\(\overline{MS}\)) in which \(\gamma_E - \log(4 \pi) - 1\) is
subtracted along with the pole~\cite{GL84}.

The result for \(\delta M_{loop}^2\) is the same for all pions (as
required by the graded symmetry group) including the physical,
unquenched pions. It must thus contain discretization errors
proportional to the combination \(2w_6' + w_8'\), as we see is the
case. The result (\ref{eq:deltmloop}) agrees with that given in
Ref.~\cite{ABB}.

Non-analytic contributions to the scattering amplitudes arise 
from the diagrams of Fig.~\ref{fig:infvol}(c-f), from wavefunction renormalization, and also from mass renormalization.
The latter contribution is, at the order we work,
only present in $\CS$, and arises kinematically from the LO result.
To see this explicitly note that the general, off-shell form of the LO amplitude is
\begin{equation}
\mathcal{S}^{(0)}_{\mathrm{off}}(s,t,u) = w_8' + \frac{1}{6f^2}
\left(2 M_0^2 - 2s + t + u \right) \,.
\label{eq:soff}
\end{equation}
When going on shell one sets $p^2=k^2=p'^2=k'^2$ to $-M_\pi^2$, with
$M_\pi$ the pion mass at the order being calculated.
This implies that $s+t+u=4(M_0^2+\delta M^2)$.
Substituting this in (\ref{eq:soff}) one finds a NNLO
contribution to $\CS$ of
\begin{equation}
\mathcal{S}^{(2)}_{\delta M^2}= \frac{2 \delta M^2}{3f^2}
\,.
\end{equation}
Combining this with the other loop contributions we find that the entire one-loop contribution to the PQ amplitudes is
\begin{equation}
\begin{split}
\label{eq:dloop}
\mathcal{D}^{(2)}_{loop}(s,t,u) & = - 2 \delta z_{loop}
\mathcal{D}^{(0)} + \frac{2 M_0^2}{9f^2} I_1 \\ & + (1/36)
\left[I_6(s)+I_6(t)+I_6(u)\right] \\ & + \left[ -(10/3) w_6' - 2 w_8'
+ 2 w_7'\right] I_1 \\ & - (4/3) w_6' I_4(t) \\ & + (1/3) \tilde w_8'
\left[ I_4(s) + I_4(u) - 2 I_4(t)\right] \\ & + w_6'^2 \left[4 I_2(s)
+ 4 I_2(u) + 14 I_2(t)\right] \\ & + 10 w_6' \tilde w_8' I_2(t) \\ & +
\tilde w_8'^2 \left[ I_2(s) + I_2(u) + (17/2) I_2(t)\right]
\end{split}
\end{equation}
and
\begin{equation}
\begin{split}
\label{eq:sloop}
\mathcal{S}^{(2)}_{loop}(s,t,u) & = \frac{2}{3f^2} \delta M_{loop}^2 -
2\delta z_{loop} \mathcal{S}^{(0)}(s) \\ & + \frac{10 s -
13M_0^2}{18f^2} I_1 \\ & + (1/18) \left[I_7(t,u)+I_7(u,t)\right]
\\ & - \left[(11/3) w_6' + (3/2) w_8' + 2 w_7'\right] I_1 \\ & + (1/3)
w_6' \left[ 2 I_4(s) - I_4(t) - I_4(u) \right] \\ & - (1/3) \tilde
w_8' \left[I_4(t) + I_4(u)\right] \\ & + 4 w_6' \tilde w_8'
\left[I_2(s)+I_2(t)+I_2(u)\right] \\ & - (5/2) \tilde w_8'^2 \left[
I_2(t)+I_2(u)\right] \,,
\end{split}
\end{equation}
where
\begin{equation}
\tilde w_8' = w_8' + \frac{M_0^2}{3f^2} \,.
\end{equation}
In each case, the results proportional to \(I_1\) are from 
the tadpole diagram, Fig.~\ref{fig:infvol}(c).

The new integrals that appear, and their values
after subtraction and going on-shell, are
\begin{align}
\label{eq:i2def}
\begin{split}
I_2(s) & = \int_q \frac{1}{(q^2+M_0^2)(\tilde q^2+M_0^2)} \\ &
\rightarrow \frac{1}{16\pi^2}\left[F(s) + 1 - \log(M_0^2/\mu^2)\right]
\end{split} \\
\label{eq:i4def}
\begin{split}
I_4(s) & = \frac1{f^2}\int_q \frac{-5s+t+u-2q^2-2\tilde
q^2}{2(q^2+M_0^2)(\tilde q^2+M_0^2)} \\ & \rightarrow I_2(s)\frac{4
M_0^2-3s}{f^2} - 2 I_1
\end{split} \\
\label{eq:i6def}
\begin{split}
I_6(s) & = \frac1{f^4} \int_q (3s + p^2 + k^2 + q^2 + \tilde q^2)\\ &
\times \frac{(3s + p'^2 + k'^2 + q^2 + \tilde q^2)}{(q^2+M_0^2)(\tilde
q^2+M_0^2)}\\ & \rightarrow I_2(s) \frac{(3 s-4 M_0^2)^2}{f^4} + I_1
\frac{10 s-16 M_0^2}{f^2}
\end{split}\\
\label{eq:i7def}
\begin{split}
I_7(s,u) & = \frac1{f^4} \int_q (3(p + q)^2 - p^2 - k^2 - q^2 -
\tilde q^2)\\ & \times \frac{(3(k' + q)^2 - p'^2 - k'^2 - q^2 - \tilde
q^2)}{(q^2+M_0^2)(\tilde q^2+M_0^2)}\\ & \hspace{-50pt} \rightarrow
I_2(s) \frac{3 s^2 + 3 us/2 - 12 M_0^2 s - 6M_0^2 u + 16 M_0^4}{f^4}\\
& \hspace{-50pt} + I_1 \frac{4s + 3 u - 10 M_0^2}{f^2} + \frac{(6
M_0^2 - s)(4 M_0^2 - s - 2u)}{32 \pi^2 f^4} \,,
\end{split}
\end{align}
where \(\tilde q = q + p + k\).
Thus all integrals can be written in terms of $I_1$ and 
a function defined in Ref.~\cite{ABB}:
\begin{equation}
F(s) = - \sigma_s \log \frac{\sigma_s + 1}{\sigma_s-1},
\ \ \sigma_s = \sqrt{1 - \frac{4 M_0^2}{s}}\,.
\end{equation}

The analytic contributions of NLO and NNLO arise from 
Figs.~\ref{fig:infvol}(g) and (h).
For the mass shift and wavefunction renormalization we find
\begin{align}
\begin{split}
\delta M_{an}^2 & = -(1/2)(2 w_4 + w_5)M_0^2 + (2 w_6 + w_8)\chi
\\ & {} -(1/2)(2 \xi_4 + \xi_5) M_0^2 + (2 \xi_6 +\xi_8)\chi
\\ & {} + \delta M_{ad}^2
\end{split}\\
\delta z_{an} & = (1/2)\left[(2 w_4 + w_5) + (2 \xi_4 +
\xi_5)\right] + \delta z_{ad} \,,
\end{align}
where
\begin{equation}
w_k = \frac{16 \hat a W_k}{f^2} 
\qquad \xi_k = \frac{16 L_k \chi}{f^2}
\qquad (k=4-8)
\end{equation}
are defined in analogy with Eq.~(\ref{eq:lilwdef}).
\(\delta M_{ad}^2\) and \(\delta z_{ad}\) are the
contributions from the ``additional'' terms, 
meaning \(a^3\), \(a^2 m\), \(a^4\) and \(a^2 p^2\) terms. 
They are given in Appendix \ref{app:indcon}.

Incorporating the corrections from \(\delta M_{an}^2\) and \(\delta
z_{an}\) we determine the NLO and NNLO analytic terms in the PQ
amplitudes to have the explicit form
\begin{align}
\label{eq:dan}
\begin{split}
\mathcal{D}^{(1,2)}_{an}(s,t,u) 
   & = (2 w_4 + w_5)(-2 w'_6) + w_4 G(t)
\\ & + 2 \chi w_6/f^2 
\\ & + 8 L_1 G(t)^2 + 4 L_2 \left(G(s)^2 + G(u)^2\right) 
\\ & - 4 L_{PQ} \left(G(s)^2 + G(t)^2 + G(u)^2\right) 
\\ & + (2 \xi_4 + \xi_5) (-2 w_6') + \xi_4 G(t) 
\\ & + 2 \chi \xi_6 /f^2 + \mathcal{D}_{ad}(t)\,,
\end{split}\\
\label{eq:san}
\begin{split}
\mathcal{S}^{(1,2)}_{an}(s,t,u) 
   & = (2 w_4 + w_5)(-w_8' - M_0^2/f^2)
\\ & + w_4 s/(2f^2)\\ & + 2 \chi (w_6 + w_8)/f^2 
\\ & + 2 L_3 \left(G(t)^2 + G(u)^2\right)
\\ & + 4 L_{PQ} \left(G(s)^2 + G(t)^2 + G(u)^2\right) 
\\ & + (2\xi_4+\xi_5)(-w_8' - M_0^2/f^2) 
\\ & + \xi_4 s /(2f^2) 
\\ & + 2 \chi (\xi_6+\xi_8)/f^2 + \mathcal{S}_{ad}(s) \,,
\end{split}
\end{align}
where
\begin{equation}
G(s) = \frac{s - 2 M_0^2}{f^2}
\end{equation}
and \(\mathcal D_{ad}(t)\) and \(\mathcal S_{ad}(s)\) are given in
Appendix \ref{app:indcon}. Note that, due to the great number of LECs, 
the analytic parts of the PQ amplitudes are poorly constrained.

The final results are obtained by combining Eqs.~(\ref{eq:dtree}),
(\ref{eq:dloop}) and (\ref{eq:dan}) for
\begin{equation}
\CD(s,t,u) = \CD^{(0)} + \CD^{(2)}_{loop}(s,t,u) +
\CD^{(1,2)}_{an}(s,t,u),
\end{equation}
and similarly combining Eqs.~(\ref{eq:stree}), (\ref{eq:sloop}) and
(\ref{eq:san}) for
\begin{equation}
\CS(s,t,u) = \CS^{(0)}(s) + \CS^{(2)}_{loop}(s,t,u) +
\CS^{(1,2)}_{an}(s,t,u).
\end{equation}
These may then be used in Eq.~(\ref{eq:adef}) to find 
\(\mathcal A\). We find complete agreement with the result of
Ref.~\cite{ABB}. An interesting aspect of this comparison is
that the terms linear in $a$ in $\CD$ and $\CS$
(which have the generic form $a M_0^2$ and $a s$) cancel in $\CA$. This is only true when the leading order amplitude \(\CA^{(0)}\) is expressed in terms of \(M_0\) as in Eq~(\ref{eq:atree}).

\section{Finite Volume Correlation Functions\label{sec:finvol}}

In this section we calculate, in PQWChPT, the correlation functions
appearing in the ratios shown schematically in Fig.~\ref{fig:ratios}.
Specifically, we consider a cubic spatial box 
of length \(L\) with periodic boundary conditions, and
assume that the time direction satisfies \(L_t \gg L\) and is
effectively infinite. 
Since the PQ theory does
not have a positive transfer matrix, correlation functions
cannot be analyzed by inserting complete sets of states with positive
norm.\footnote{%
As seen explicitly in the transfer matrix obtained
in Ref.~\cite{BGPQtransfer}.}
Furthermore, the scattering amplitudes of the infinite volume
theory are not unitary. Because of these unphysical features, the
standard relation due to L\"uscher between finite volume energy shifts
and phase shifts~\cite{Luscher:1986pf} 
does not hold---neither energy shifts nor phase shifts can be defined.
Instead, if one wants to ``measure''
scattering amplitudes, one can compare results for 
Euclidean correlation functions (obtained using lattice methods)
with the predictions of PQWChPT. 
Then the unphysical features of the underlying theory are
matched by those of the effective theory. This strategy was introduced
in Ref.~\cite{BGpipi} to study pion scattering in the quenched
approximation.

To approach infinite-volume scattering as closely as possible, we
consider external pion fields with definite three-momentum, leaving
only their time coordinates untransformed.
We additionally restrict ourselves to the simplest case of pions at rest. 
Specifically, we calculate the following four-point correlators:
\begin{align}
\label{eq:cd}
C_{\mathcal{D}}(t) & = 
\langle \tilde \pi_{12}(t) \tilde \pi_{34}(t) 
            \tilde \pi_{21}(0) \tilde \pi_{43}(0) \rangle \\
\label{eq:cs}
C_{\mathcal{S}}(t) & = 
\langle \tilde \pi_{12}(t) \tilde \pi_{34}(t) 
            \tilde \pi_{23}(0) \tilde \pi_{41}(0) \rangle  \,,
\end{align}
with
\begin{equation}
\tilde \pi(t) = \int_{L^3} d^3 x\, \pi(\vec x, t)  \,.
\label{eq:pitildedef}
\end{equation}
The subscript on the integral indicates integration over finite volume. 
These correlation functions are, roughly speaking,
the finite-volume analogs of the infinite-volume 
unamputated scattering amplitudes.
In order to more directly access these amplitudes,
we take the ratio of these correlators 
to the square of a single-pion correlator,
which is, roughly speaking, the analog of amputation:
\begin{align}
\label{eq:Rdef} 
R_{\mathcal{D},\mathcal{S}}(t) & = 
\frac{C_{\mathcal{D},\mathcal{S}}(t)}{C_\pi(t)^2} \\
\label{eq:Cpidef}
C_\pi(t) & = \langle \tilde\pi_{12} (t) \tilde \pi_{21}(0)\rangle  \,.
\end{align}
These are the ratios shown schematically in Fig.~\ref{fig:ratios}.
In this note we will always take $t$ to be positive.

It is instructive to make contact with the more 
familiar results in a physical (i.e. unquenched) theory.
Adapting Eq.~(\ref{eq:apiplus}) to finite volume, 
we find that the \(\pi^{+}\pi^+\) correlation function
is related to our PQ correlators in a simple way:
\begin{equation}
\begin{split}
\label{eq:Cpipidef}
C_{\pi^+}(t) & = 
\langle \tilde \pi^+(t) \tilde \pi^+(t) 
\tilde \pi^-(0) \tilde \pi^-(0)\rangle \\
& =  2 C_\mathcal{D}(t) + 2 C_\mathcal{S}(t)  \,.
\end{split}
\end{equation}
It follows that
\begin{equation}
\label{eq:transfer}
R_\mathcal{D}(t)+R_\mathcal{S}(t) = 
Z e^{-(E_{\pi^+\pi^+}-2 M_\pi) t} + {\rm exp.\ supp.,}
\end{equation}
where \(E_{\pi^+\pi^+}\) is the energy of the lightest \(\pi^+\pi^+\)
state with zero total spatial momentum, and the exponentially
suppressed terms come from states with higher energies.
At LO in WChPT one finds (as we will show below) that
the overlap factor \(Z\) is unity, so that
\begin{equation}
\begin{split}
R_\CD(t)+R_\CS(t) & \approx 1 - \delta E t \\
\delta E & =  E_{\pi^+\pi^+}-2 M_\pi  \,.
\end{split}
\end{equation}
This approximation is valid as long as \(t\) is large enough that the
exponentially suppressed terms are negligible but also small enough
that the linear term dominates the Taylor expansion of the leading
exponential. 

L\"uscher's result relates this shift to the infinite volume 
$\pi^+$ scattering amplitude at threshold~\cite{Luscher:1986pf}:
\begin{eqnarray}
\delta E &=& - {\cal A}_{\pi^+}^{\rm th}\frac{1}{4 M_\pi^2 L^3} 
\left[1 + c_1 {\cal A}_{\pi^+}^{\rm th} \frac{1}{16\pi M_\pi L}
+ O(1/L^2) \right] 
\nonumber\\
&& 
+ \CO(e^{-M_\pi L}) \,,
\label{eq:Luscher}
\end{eqnarray}
where \(c_1=-2.837297\) is a numerical constant. 
The form of the \(1/L^2\) correction, and some higher order terms, are
known~\cite{Luscher:1986pf}, 
but we do not show them as we will not control the
corresponding terms in our calculation. 
Note also the presence of exponentially suppressed finite-volume
corrections. 

Although the PQ ratios \(R_\CD(t)\) and \(R_\CS(t)\) do not behave as
a sum of exponentials, what we can take over from the analysis of
\(R_{\pi^+}(t)\) is that it is useful to determine the coefficient of
the term linear in \(t\). In the remainder of this section we
determine the PQWChPT prediction for the linear terms in \(R_\CD(t)\)
and \(R_\CS(t)\). We work to NNLO in the momentum power counting of
Eq.~(\ref{eq:powcou}) and do so controlling not only the leading
\(1/L^3\) terms but also the \(1/L^4\) corrections. We can also
control a subset of the finite volume corrections proportional to
\(e^{-M_\pi L}\), but will not do so systematically. 

The PQWChPT diagrams which contribute to the order we 
work are shown in Fig.~\ref{fig:fig1}. 
These are the same diagrams as for the infinite-volume amplitudes,
Fig.~\ref{fig:infvol}, except for the addition of the disconnected
diagrams (a-c).
Our description of the calculation is broken into three subsections: leading-order results, analytic NLO and NNLO contributions, and NNLO results from loop diagrams.

\subsection{Leading-order Results}
\label{subsec:LO}

\begin{figure}[tbp!]
{\includegraphics[angle=270,width=0.45\textwidth]{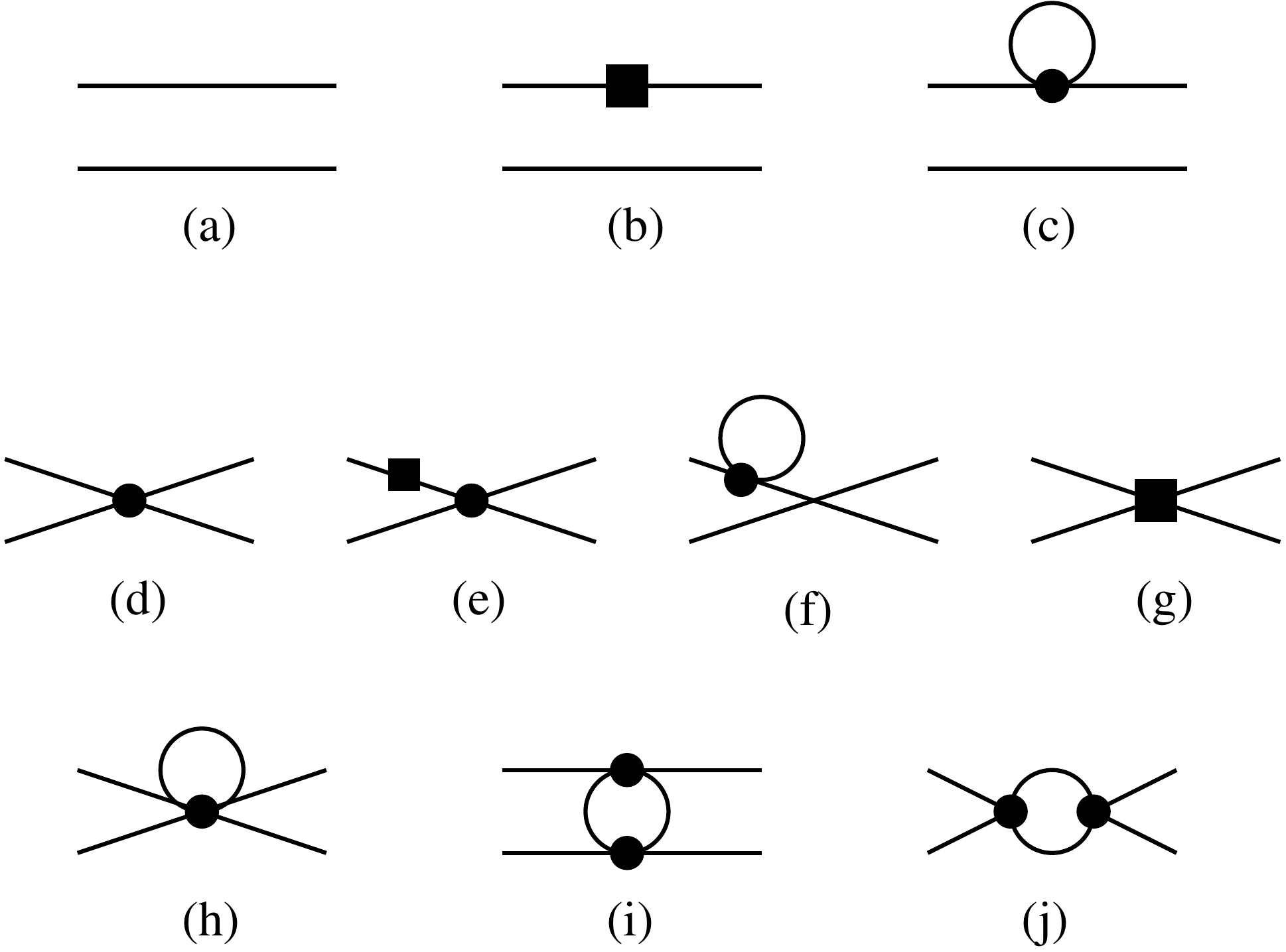}}
\caption{Classes of diagrams contributing to the correlators
\(C_\CD(t)\) and \(C_\CS(t)\) in ChPT. 
Diagrams in which initial or final pions are interchanged 
are not shown separately. 
Notation for vertices as in Fig.~\ref{fig:infvol}.
\label{fig:fig1}}
\end{figure}

The pion-disconnected diagram of
Fig.~\ref{fig:fig1}(a) contributes only to \(C_\CD(t)\), with the
result
\begin{gather}
C_\CD^{({\rm disc})}(t) = C_\pi(t)^2 \\
C_\pi(t)
= \frac{L^3}{2 M_0} e^{-M_0 t}
\end{gather}
and thus
\begin{equation}
R_\CD^{({\rm disc})}(t) = 1\,,\quad
R_\CS^{({\rm disc})}(t) = 0\,.\quad
\label{eq:Rdisc}
\end{equation}
The factor of \(L^3\) in \(C_\pi(t)\) arises because both ends of the
propagators are integrated over space. Note that, within WChPT, there are
no explicit contributions from excited pions at any order.
The effect of these states appears 
through contact terms proportional to \(\delta(t)\), which arise
first at NNLO.
Since we always consider \(t\) large enough to remove terms which
are exponentially suppressed, we need not include such contact terms.  

The tree-level, pion-connected diagram, Fig.~\ref{fig:fig1}(d),
contributes to both \(C_\CD\) and \(C_\CS\).
Because the four pion sources have \(\vec p=0\), only pions
at rest contribute. Thus the pion propagators attaching to the vertex
are either 
\begin{equation}
\frac{e^{-M_0 |t_1|}}{2M_0} \ \ {\rm or} \ \
\frac{e^{-M_0|t-t_1|}}{2M_0}\,,
\label{eq:propform}
\end{equation}
where $t_1$ is the time of the vertex.
These propagators have no factors of \(L^3\) 
since only one end is integrated over space. 
There is, however, a factor of \(L^3\) from integrating
the position of the vertex over space.

The correlator \(C_\CD(t)\) is the simplest to consider, because the
LO vertex comes only from the $W'_6$ term in $\CL_{\rm LO}$
[see Eq.~(\ref{eq:lolag})] and therefore contains no derivatives.
Based on the discussion above, we find (recalling that \(t\) is positive)
\begin{align}
C_\CD^{({\rm conn},0)}(t) & = 2 w_6' L^3 \int dt_1 \frac{e^{-2M_0
    |t_1|} e^{-2M_0 |t-t_1|}}{(2 M_0)^4} \\ & = 2 w_6' L^3
\frac{e^{-2M_0 t}}{(2 M_0)^4} \left(\frac{1}{2 M_0} +t\right) \,.
\end{align}
The factor of \(t\) arises from the region \(0 \le t_1 \le t\) in
which the vertex lies between the two sources.
The \(t\)-independent terms
come from \(t_1 < 0\) and \(t_1 > t\), where the contribution drops
exponentially. 

At this point, it is useful to incorporate a result from the
calculation of subleading orders.
The diagrams which correct pion propagators,
Figs.~\ref{fig:fig1}(b), (c), (e) and (f),
form part of the geometric series which changes
\begin{equation}
\frac{e^{-M_0 |t|}}{2 M_0} \ \ {\rm to}\ \
Z_\pi \frac{e^{-M_\pi |t|}}{2 M_\pi}
\,.
\end{equation}
Here $M_\pi$ is the physical pion mass to the order we are working,
and $Z_\pi= 1-\delta z$ is the wavefunction renormalization.
For the moment we incorporate only the mass-shift,
returning to the effect of $Z_\pi\ne1$ below.
Using the new propagators in both numerator and denominator 
of the ratio \(R_\CD(t)\), we find
\begin{eqnarray}
R_\CD^{({\rm conn},0)}(t) &=&  
\frac{1}{4 M_\pi^2 L^3}\left(\frac1{2 M_\pi} + t\right) 2 w_6'  \,,
\label{eq:RDconn0}
\\
&=&
\frac{1}{4 M_\pi^2 L^3}\left(\frac1{2 M_\pi} + t\right) \CD^{(0)} 
\,.
\label{eq:RDconn0gen}
\end{eqnarray}
The utility of the ratio is that the overall exponentials cancel---as
noted above, this corresponds to amputation in a continuum calculation.
The physical interpretation of the \(t\) term is that the pions can
interact at any intermediate time, while the \(1/L^3\) suppression
arises because the zero-momentum pions must overlap in order to
interact. As shown in the second line, the coefficient of $t$ is, 
aside from the kinematical factor $1/(4 M_\pi^2 L^3)$,
the LO PQ scattering amplitude. This, together with the
disconnected contribution from (\ref{eq:Rdisc}), gives
the result (\ref{eq:RDLO}) quoted in the introduction.

A similar analysis holds for \(R_\CS(t)\), except that
we must now deal with the momentum dependence
arising from the kinetic term in the LO Lagrangian (\ref{eq:lolag}).
If one evaluates the diagram in position space,
the derivatives in the vertex act on the pion propagators of
Eq.~(\ref{eq:propform}) (with $M_0\to M_\pi$). Thus only
time derivatives contribute, and they yield $\pm M_\pi$.
Consider first $0<t_1<t$, i.e. the vertex lying between the sources. 
Derivatives acting on pion propagators
originating at times $0$ and $t$ then give $-M_\pi$ and $+M_\pi$,
respectively. This implies that $s=4 M_\pi^2$ and $t=u=0$, 
i.e. on-shell kinematics at threshold.
For $t_1<0$ ($t_1>0$), by contrast, all derivatives give $+M_\pi$ ($-M_\pi$),
and one obtains, in both cases, the amplitude at off-shell kinematics:
$s=t=u=4M_\pi^2$.
The final result is
\begin{eqnarray}
R_\CS^{({\rm conn},0)}(t) &=& 
\frac{1}{4 M_\pi^2 L^3} \Bigg[
\frac{(w_8' + {M_0^2}/{3 f^2})}{2 M_\pi} 
\nonumber \\ 
&& \quad + \left(w_8' + \frac{M_0^2-4 M_\pi^2}{3 f^2}\right) t \Bigg] \,,
\label{eq:RSconn0}
\\
&=&
\frac{1}{4 M_\pi^2 L^3} \Bigg[
\frac{1}{2 M_\pi}\CS^{(0)}_{\rm off}(4M_\pi^2,4M_\pi^2,4M_\pi^2)
\nonumber \\
&&\quad + \CS^{(0)}(4M_\pi^2) t \Bigg]
\,.
\label{eq:RSconn0gen}
\end{eqnarray}
This result, together with the vanishing of the disconnected
contribution, is reported in Eq.~(\ref{eq:RSLO}) of the introduction.
Note that, since in (\ref{eq:RSLO}) we are quoting a LO result,
we can set $M_0=M_\pi$.

As a check on our results we consider the sum 
\(R_\CD(t) + R_\CS(t)\).
According to the discussion above,
this should be $\approx (1 - \delta E t)$,
with $\delta E= -\CA_{\pi^+}^{\rm th}/(4 M_\pi^2 L^2)$ 
[the LO term in Eq.~(\ref{eq:Luscher})]. 
We find (setting $M_0=M_\pi$)
\begin{multline}
R_\CD^{(\rm disc)}(t)+R_\CD^{({\rm conn},0)}(t) + 
R_\CS^{({\rm conn},0)}(t) = 
\\ \left\{1 + \CO\left(\frac1{L^3}\right) 
+ \frac{1}{4M_\pi^2 L^3}
\left( 2w_6' + w_8' - \frac{M_\pi^2}{f^2} \right) t \right\}
\,.
\label{eq:RD+RSlinear}
\end{multline}
The coefficient of $t/(4 M_\pi^2 L^3)$ is indeed the
$\pi^+$ scattering amplitude at threshold.
The $1/L^3$ corrections to the $t$-independent terms
(which are not shown in detail) are the first
correction to the $Z$-factor for the two pion state.

\subsection{Analytic NLO and NNLO contributions}
\label{subsec:analytic}

Analytic NLO and NNLO contributions
arise from Figs.~\ref{fig:fig1}(b), (e) and (g).
Diagrams (b) and (e) 
give the analytic parts of mass and wavefunction renormalization,
contributions which are identical to those in infinite volume. 
The effect of mass renormalization has already been discussed above.
Wavefunction renormalization partly cancels in the ratios,
leaving a factor of $Z_\pi^2=(1-2 \delta z)$, exactly what is
needed to renormalize the amplitude as in infinite volume.

The analytic contributions to the vertex,
Fig.~\ref{fig:fig1}(g), can be analyzed by a straightforward
generalization of the method used for the momentum-dependent
contribution to the LO vertex in $R_{\CS}$.
We find 
\begin{eqnarray}
R_{\CD,an}^{(1,2)}(t) &=& \mathrm{const} 
+ \frac{t}{4 M_\pi^2 L^3} \mathcal D_{an}^{(1,2)}(4 M_\pi^2,0,0) \,,
\label{eq:RDconnan}\\
R_{\CS,an}^{(1,2)}(t) &=& \mathrm{const} 
+ \frac{t}{4 M_\pi^2 L^3} \mathcal S_{an}^{(1,2)}(4 M_\pi^2,0,0) \,,
\label{eq:RSconnan}
\end{eqnarray}
where $\CD_{an}^{(1,2)}$ and $\CS_{an}^{(1,2)}$ are the 
infinite-volume analytic contributions to the two amplitudes given
in Eqs.~(\ref{eq:dan}) and (\ref{eq:san}). We note that, at threshold, these amplitudes are just linear combinations of \(a^3\), \(a M_\pi^2\), \(a^4\), \(a^2 M_\pi^2\) and \(M_\pi^4\) with independent coefficients.
The results (\ref{eq:RDconnan}) and (\ref{eq:RSconnan}) are the
natural generalizations of Eqs.~(\ref{eq:RDconn0gen})
and (\ref{eq:RSconn0gen}).
Note that we only keep track of the terms linear
in $t$, since these are proportional to the desired PQ
amplitudes at threshold.
The constant terms involve off-shell amplitudes. 

\subsection{NNLO results from loop diagrams}
\label{subsec:loops}

In this section we extend the calculation to the one loop diagrams
which appear at NNLO, focusing on the coefficient of the terms linear
and quadratic 
in \(t\) in the ratios \(R_{\CD,\CS}(t)\). At one loop there are 3
types of contributions, which we discuss in order of increasing
complexity. First, there are
tadpole diagrams, shown in Fig.~\ref{fig:fig1}(c), (f) and (h).
Second there are \(t\) and \(u\)-channel loops, exemplified by
Fig.~\ref{fig:fig1}(i). And, third, there are \(s\)-channel loops, as
shown in Fig.~\ref{fig:fig1}(j). 

\subsubsection{Tadpole diagrams}
\label{subsubsec:tadpole}

Tadpole diagrams on external legs [Figs.~\ref{fig:fig1}(c) and (f)]
renormalize the external pion propagators,
contributing to $\delta M^2$ and $\delta z$ as described
for the analytic terms.
In this case, however, there is a difference compared to infinite volume, 
namely that
in the tadpole integral \(I_1\) one should use the finite volume pion
propagator. This gives rise to corrections which are suppressed by
powers of \(\exp(-M_\pi L)\), as can be seen by implementing the
periodic boundary conditions using images.
We assume such corrections are negligible.

Tadpole diagrams attached to the vertex [Fig.~\ref{fig:fig1}(h)] are
also simple to incorporate. As long as \(0 < t_1 < t\), they
multiply the tree-level vertex by a factor that is independent of the
vertex position, leading to the usual term linear in $t$.
The coefficient of $t$ is proportional to the tadpole contributions
to the threshold amplitudes, and is exactly that needed to maintain the
forms of Eqs.~(\ref{eq:RDconnan}) and (\ref{eq:RSconnan}),
with $\CD$ and $\CS$ now including one-loop tadpole contributions.
Again, there are exponentially suppressed volume corrections which
we assume negligible.

\begin{figure}[tbp!]
{\includegraphics[angle=270,width=0.45\textwidth]{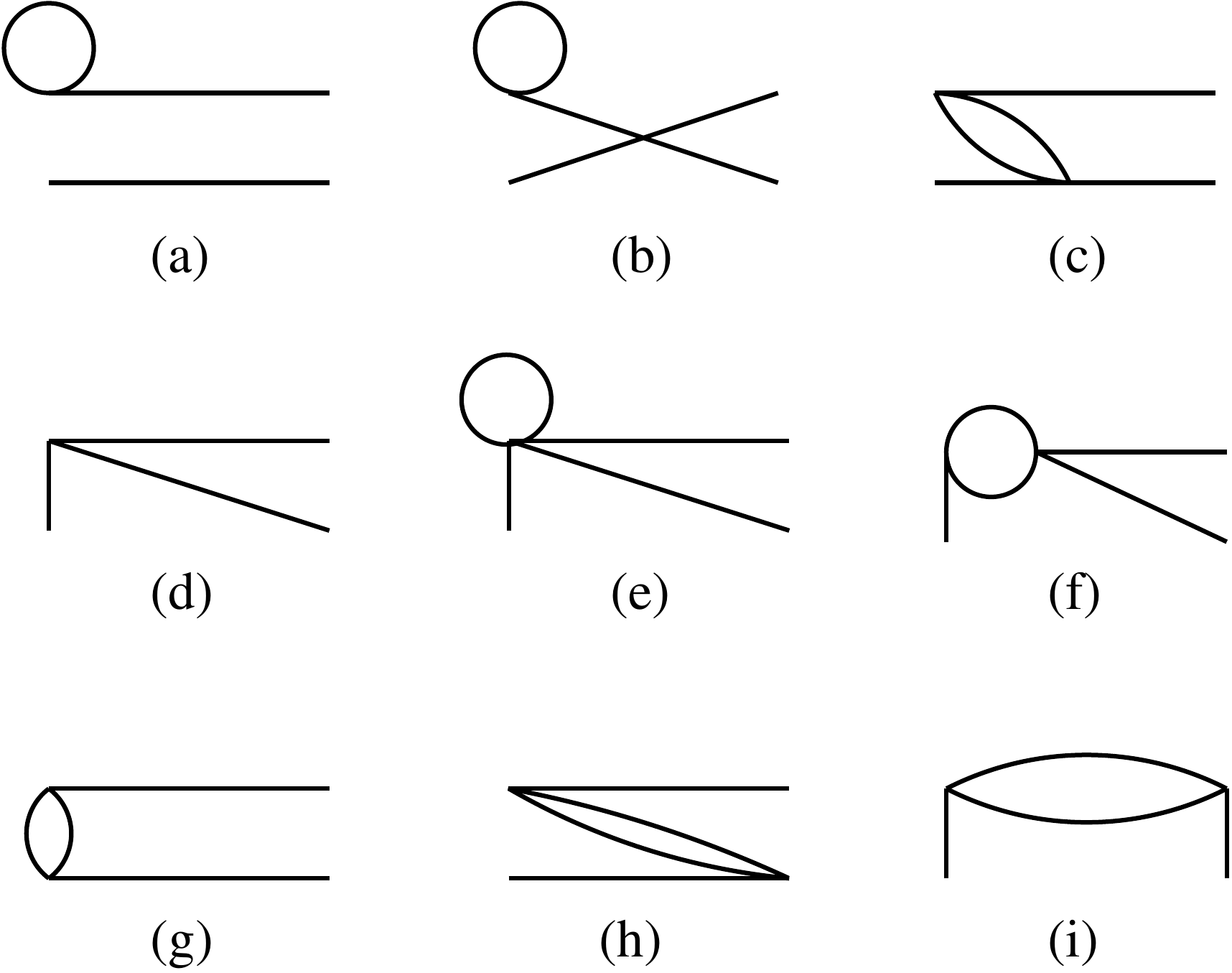}}
\caption{Classes of diagrams contributing to the correlators
\(C_{\CD}(t)\) and \(C_{\CS}(t)\),
 arising from \(\pi^3\) and \(\pi^5\) terms in
the external operators. All vertices are from the LO Lagrangian.
\label{fig:fig2}}
\end{figure}

At this stage, it is appropriate to mention that there can also be
tadpoles arising on the external operators at times \(0\) and
\(t\). This is because, in practice, if one uses a local pseudoscalar
operator at the quark level, then it maps into the chiral theory 
at LO as \(\bar q_k \gamma_5 q_j(x) \to c
(\Sigma-\Sigma^\dagger)_{jk}(x)\). The constant \(c\) is known but
unimportant here, since it cancels in the ratios. This chiral operator
expands to a term proportional to \(\pi_{jk}\)---the operator we have
been using---but with corrections proportional to \((\pi^3)_{jk}/f^2\)
and \((\pi^5)_{jk}/f^4\). These corrections give rise, at the order we
are working, to the diagrams of Fig.~\ref{fig:fig2}.
As we now explain, however, all of these diagrams lead to contributions
subleading compared to those we are keeping.

Tadpoles associated with external operators [Figs.~\ref{fig:fig2}(a) and (b)]
cancel in the ratios. The loop diagram of Fig.~\ref{fig:fig2}(c) does
not give rise to a term linear in \(t\) because the four propagators
to the left of the vertex lead to a fall off of at least \(e^{-4M_\pi
  t_1}\). The tree-level diagram of Fig.~\ref{fig:fig2}(d) also does not
contribute to the coefficient of \(t\)---it gives a \(1/L^3\)
contribution to the constant term. The same is true for the
corresponding tadpole diagram, Fig.~\ref{fig:fig2}(e). The only
diagram in this class that does give a contribution linear in \(t\) is
Fig.~\ref{fig:fig2}(f). This arises when both pions in the loop have
\(\vec q=0\). This contribution is, however, of size \(t/L^6\) in the
ratios,\footnote{%
The \(1/L^6\) occurs because the numerator in the ratios is
independent of \(L\), while the denominator is proportional to
\(L^6\). The numerator is independent of \(L\) because the three
external propagators have one leg summed and are thus
\(L\)-independent, leaving two point-to-point propagators in the loop
(each proportional to \(1/L^3\)) and two vertices integrated over space
(each giving a factor of \(L^3\)).}
and thus is subleading to the \(1/L^4\) terms that we are aiming to
control.

There are also analytic corrections to the external operators, e.g.
terms proportional to $\partial^2 \pi_{jk}$ and $a^2 \pi_{jk}$.
These lead to corrections which cancel in the ratios.

\subsubsection{\(t\) and \(u\)-channel loops}
\label{subsubsec:t-u}

The \(t\) and \(u\)-channel loop diagrams have the form of
Fig.~\ref{fig:fig1}(i). Note that, since both external pions at time
\(t\) are summed over all space, there is no difference between the $t$
and \(u\)-channel loops. In order to avoid confusion between the two uses
of \(t\) we will couch our discussion in terms of the \(u\)-channel.

Figure~\ref{fig:fig1}(i)
gives rise to a contribution proportional to the time
separation \(t\) as follows. Although there are two vertices (at times
\(t_1\) and \(t_2\)), when they are pulled apart in Euclidean space
there is an exponential suppression, so the dominant contributions
occur when \(|t_1-t_2|< 1/M_\pi\). The loop collapses to an
effective vertex, which, when integrated over the
intermediate time, leads to a factor of \(t\). If either of the
vertices is outside of the region \(0< t_{1,2}<t\), it is easy to see
that one does not get a term linear in \(t\).

We now turn these words into a concrete evaluation. We consider first
a contribution in which both vertices have no derivatives. In infinite
volume, this leads to those terms in Eqs.~(\ref{eq:dloop}) and
(\ref{eq:sloop}) which contain the integral \(I_2(u)\). For
definiteness, we consider the contribution to 
\(\CD^{(2)}_{loop}\) of \(4 w_6'^2 I_2(u)\). 
The corresponding contribution to the finite-volume correlator is
\begin{align}
C_\CD(t) & \supset \frac{e^{-2M_\pi t}}{(2M_\pi)^4} 4 w_6'^2 L^3
\tilde I_2(t)\\ \tilde I_2(t) & = \frac{1}{L^3} \int_0^t dt_1 \int_0^t
dt_2 \int_{\vec x_1} \int_{\vec x_2} G_\pi(x_1,x_2)^2 \,.
\end{align}
Here \(G_\pi(x_1,x_2)\) is the Euclidean pion propagator, which is
related to \(C_\pi(t)\) by
\begin{equation}
C_\pi(t_1-t_2) = \int_{\vec x_1} \int_{\vec x_2} G_\pi(\vec
x_1,t_1;\vec x_2,t_2) \,.
\end{equation}
The contribution to the ratio is
\begin{equation}
R_\CD(t) \supset \frac1{4 M_\pi^2 L^3} 4 w_6'^2 \tilde I_2(t)
\,.
\end{equation}
It is straightforward to evaluate $\tilde I_2(t)$ explicitly, 
and one finds that it consists of a term linear in $t$ up to
corrections falling as $e^{-2 M_\pi t}$.
A simple way to pick out the coefficient of $t$
is to take a time derivative and
then send $t\to\infty$. In this way we arrive at
\begin{equation}
R_\CD(t) \supset \frac{t}{4 M_\pi^2 L^3} 4 w_6'^2 \tilde I_2'(\infty) 
\,,
\end{equation}
for sufficiently large $t$.

To evaluate $\tilde I_2'(\infty)$ we note
that \(\tilde I_2(t)\) can be rewritten as
\begin{equation}
\tilde I_2(t) = 2\int_0^t dt_1 \int_0^{t_1}dt_2 \int_{\vec x_2}
G_\pi(\vec 0,t_1;\vec x_2,t_2)^2 \,,
\label{eq:tloop}
\end{equation}
where we have used translation invariance and the symmetry
\(G_\pi(x,0)=G_\pi(0,x)\). Thus its derivative is
\begin{equation}
\tilde I_2'(t) = 2 \int_0^t dt_2 \int_{\vec x_2} G_\pi(\vec 0,0;\vec
x_2,t_2)^2 \,.
\label{eq:tloop2}
\end{equation}
Given the exponential fall-off of \(G_\pi\),
this integral asymptotes to its large \(t\) value once
\(t\gg 1/M_\pi\). The
factor of 2 can be traded for an extension of the integral to negative
values of \(t\). In this way we find 
\begin{align}
\tilde I_2'(\infty) & = \int_{-\infty}^{\infty} dt_2 \int_{\vec x_2}
G_\pi(\vec 0,0;\vec x_2,t_2)^2 \\ & = \int \frac{dq_4}{2\pi}
\frac{1}{L^3} \sum_{\vec q=2\pi \vec n/L} \widetilde G_\pi(q)^2 \,,
\end{align}
where \(\widetilde G_\pi\) is pion propagator in momentum space.
This is simply the finite volume version of 
\(I_2(u=0)=\int_q 1/(q^2+M_0^2)^2\):
\begin{equation}
\tilde I_2'(\infty) = I_2(u=0) + \CO\left(e^{-M_\pi L}\right) \,,
\end{equation}
where images can be used to see that the finite-volume corrections fall exponentially.

The conclusion of this analysis is that, for sufficiently large $t$ and $L$,
the diagram which leads to a contribution to $\CD^{(2)}_{loop}$
of $4 w_6'^2 I_2(u)$ contributes to the finite-volume ratio as
\begin{equation}
R_\CD(t) \supset \frac{t}{4 M_\pi^2 L^3} 4 w_6'^2 I_2(u\!=\!0) 
\,.
\end{equation}
Thus, once again, the coefficient of $t/(4 M_\pi^2 L^3)$ is the
infinite-volume scattering amplitude at threshold.
The same argument goes through identically for $I_2(u)$
contributions to $\CS^{(2)}_{loop}$,
and for contributions to both amplitudes proportional to
$I_2(t)$ ($t$ here the Mandelstam variable).

A similar argument also holds for the contributions
to $\CD^{(2)}_{loop}$ and $\CS^{(2)}_{loop}$ proportional to the
integrals \(I_4(t)\), \(I_4(u)\), \(I_6(t)\), \(I_6(u)\),
\(I_7(t,s,u)\) and \(I_7(u,s,t)\). These arise when one or both of the
vertices are from the kinetic term in the Lagrangian. 
By similar manipulations to those above, one finds that the
contribution proportional to $t$ involves exactly the
integrands of the infinite volume forms (\ref{eq:i4def}-\ref{eq:i7def}),
evaluated at threshold,
but the spatial momentum-integrals are again replaced by finite-volume sums.
The short-distance divergences are unaffected by the finiteness of the volume,
so the regularization and subtractions are unchanged.
Thus the manipulations relating these integrals to $I_2$ and $I_1$
go through, as always up to exponentially small volume corrections. 
The net effect is that the
forms of Eqs.~(\ref{eq:RDconnan}) and (\ref{eq:RSconnan}) are maintained,
with $\CD$ and $\CS$ now including \(t\) and \(u\)-channel contributions.

There is one subtlety for terms with two or more derivatives acting on the same internal propagator. To illustrate this point we define the integral
\begin{multline}
\label{eq:twoder}
\tilde I_{\partial^2}(t) = \frac{1}{L^3} \int_0^t dt_1 \int_0^t dt_2 \int_{\vec x_1} \int_{\vec x_2}  \\
\times [\partial_{x_2}^2 G_\pi(x_1;x_2)]  G_\pi(x_1;x_2) \,.
\end{multline}
Although the integrand of (\ref{eq:twoder}) is an even function of
\(x_1 - x_2\), one must be careful in carrying out the manipulations
which give the analog of Eq. (\ref{eq:tloop}). The issue is that
\(G_\pi\) has a cusp at \(t_1=t_2\) and therefore the time derivatives
give a delta-function: \(\delta(t_1-t_2)\). Carefully including the
region about the delta-function, we find
\begin{multline}
\tilde I_{\partial^2}(t) = \lim_{\epsilon \rightarrow 0} \int_0^t dt_1
\left[ 2 \int_0^{t_1-\epsilon} dt_2 +
  \int_{t_1-\epsilon}^{t_1+\epsilon} dt_2 \right] \int_{\vec x_2}
\\ \times [\partial_{x_2}^2 G_\pi(\vec 0,t_1;\vec x_2,t_2)] G_\pi(\vec
0,t_1;\vec x_2,t_2) \,.
\end{multline}
We now proceed as above, taking the time derivative and sending \(t
\rightarrow \infty\) to deduce
\begin{align}
\begin{split}
\tilde I'_{\partial^2}(\infty) & = \lim_{\epsilon \rightarrow 0} \left
       [\int_{-\infty}^\epsilon dt_2 + \int_{-\epsilon}^\epsilon dt_2
         + \int_\epsilon^\infty dt_2 \right] \\ & \hspace{50pt} \times
       \int_{\vec x_2} [\partial_{x_2}^2 G_\pi(0;x_2)] G_\pi(0;x_2)
\end{split}\\
\label{eq:dtloop}
& = \int d^4 x_2 [\partial_{x_2}^2 G_\pi(0;x_2)] G_\pi(0;x_2) \,.
\end{align}
As above, we are left with a single integral over all of space-time
(with space finite). Note however that if the middle integral were
absent then the final equality would not hold. Proceeding from
Eq.~(\ref{eq:dtloop}), it is straightforward to see that
\(I'_{\partial^2}(\infty)\) is equal to the corresponding
finite-volume amplitude at threshold.

\subsubsection{s-channel loop}
\label{subsubsec:s}

The \(s\)-channel loop diagram is shown in Fig.~\ref{fig:fig1}(j). This
diagram leads to the terms proportional to \(I_2(s)\), \(I_4(s)\) and
\(I_6(s)\) in the infinite volume amplitudes.

\begin{figure}[tbp!]
{\includegraphics[angle=270,width=0.45\textwidth]{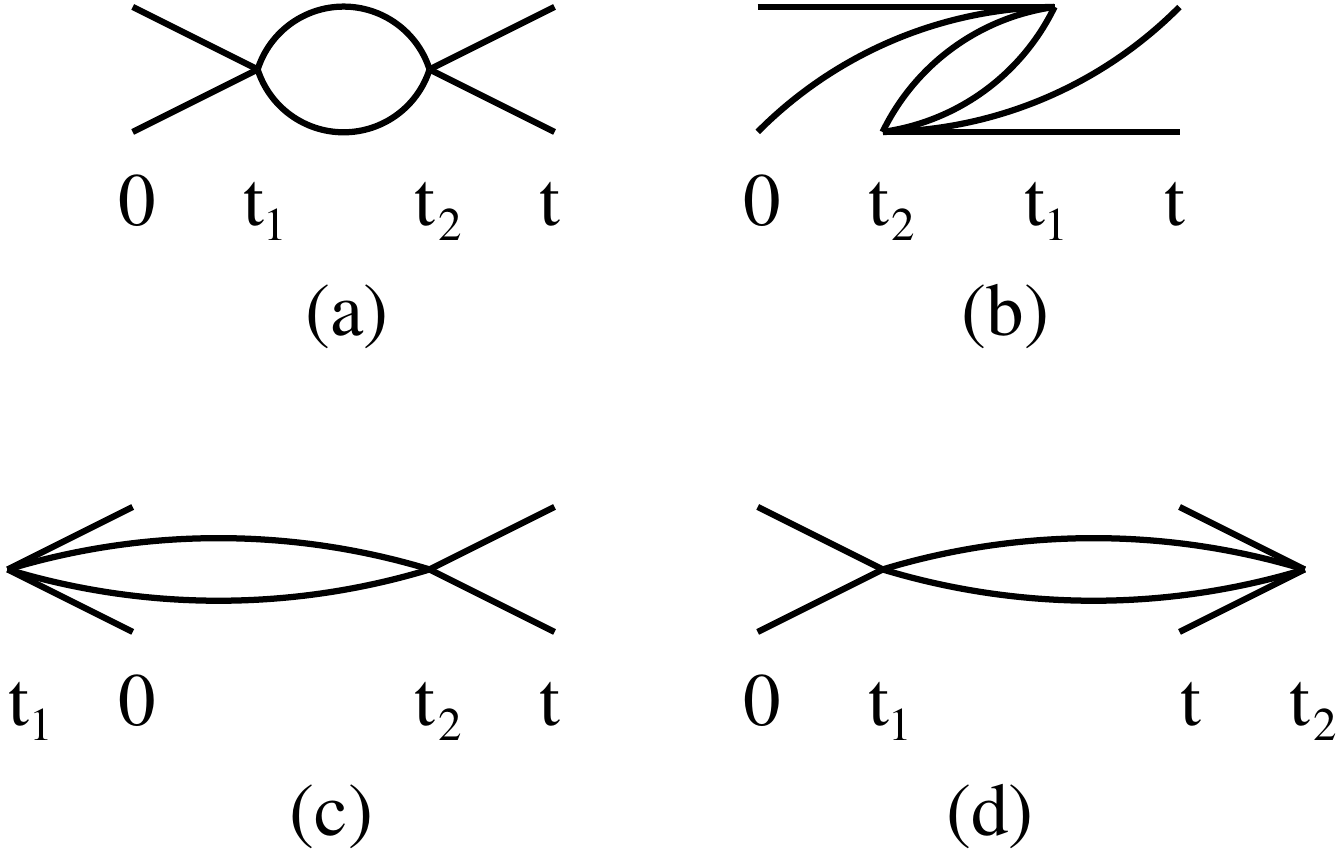}}
\caption{Time-orderings of the \(s\)-channel loop which give rise to
  contributions to \(R_{\CD,\CS}(t)\) linear in \(t\) at large times.  The
  time-ordering (a) also give rise to a quadratic term, as discussed
  in the text.
\label{fig:fig3}}
\end{figure}

We begin by analyzing the case in which neither vertex has momentum
dependence, which leads to the integral \(I_2(s)\) in infinite
volume. For definiteness we focus on the \(\tilde w_8'^2 I_2(s)\)
contribution to \(\CD^{(2)}_{loop}\) in Eq.~(\ref{eq:dloop}). 
The corresponding contribution to \(C_\CD(t)\) is
\begin{multline}
C_\CD(t) \supset \\ \tilde w_8'^2 \int_{x_1} \int_{x_2}
\frac{e^{-2M_\pi|t_1|}}{(2M_\pi)^2} \left[G_\pi(x_1,x_2)\right]^2 
\frac{e^{-2M_\pi|t-t_2|}}{(2M_\pi)^2}\,.
\label{eq:CAs}
\end{multline}
There are many choices of the ordering of the four times \(0\),
\(t_1\), \(t_2\) and \(t\), but only the four shown in
Fig.~\ref{fig:fig3} lead to terms linear in \(t\). The others give
constants or terms which fall exponentially with \(t\).

In fact, the time ordering of Fig.~\ref{fig:fig3}(a) also gives a term
quadratic in \(t\). This occurs when both pions in the loop are at
rest. The integrand is then independent of \(t_1\) and \(t_2\), so
that the contribution to the ratio is
\begin{equation}
R_\CD(t) \supset \tilde w_8'^2 \left(\frac1{4 M_\pi^2 L^3}\right)^2
\frac{t^2}2 \,.\label{eq:quadratic}
\end{equation}
In a physical theory, this is one of the contributions which builds up
the quadratic term in the expansion of \(\exp(-\delta E t)\) in
Eq.~(\ref{eq:transfer}), and we will use this below as a check on our
result. First, however, we determine the quadratic terms arising
from the finite volume versions of the integrals \(I_4(s)\) and
\(I_6(s)\).

These integrals have one or both vertices from the kinetic term in
\(\mathcal L_{LO}\) (Eq.~(\ref{eq:lolag})). For our particular
kinematics the derivatives are simple to evaluate, since all pions are
on-shell and at rest. The derivatives in the kinetic vertices thus
give exactly the results that we obtain in infinite volume when
working at threshold and on-shell (\(s=4 M_\pi^2\), \(t=u=0\)).
This means that, with these
substitutions, we can use Eqs.~(\ref{eq:i4def}) and (\ref{eq:i6def})
to relate the {\em finite volume} versions of \(I_4(s)\) and
\(I_6(s)\) to that of \(I_2(s)\). The \(I_1\) terms are to be dropped,
since they arise from the \(\vec q\ne0\) part of the loop.

Putting this together, we find the total quadratic terms to be:
\begin{align}
\label{eq:RDquad}
R_\CD(t) &\supset \left[\left(\tilde w_8' - \frac{4 M_\pi^2}{3
    f^2}\right)^2 + 4 w_6'^2\right] \left(\frac{1}{4 M_\pi^2
  L^3}\right)^2 \frac{t^2}2 \\
\label{eq:RSquad}
R_\CS(t) & \supset 4 w_6' \left(\tilde w_8' - \frac{4 M_\pi^2}{3
  f^2}\right) \left(\frac1{4 M_\pi^2 L^3}\right)^2 \frac{t^2}2 \,.
\end{align}
Thus, if one could determine the coefficients of the quadratic terms
in a simulation, one would gain additional information concerning
\(w_6'\) and \(w_8'\). Note that the coefficients of the quadratic
terms are not the squares of those of the linear terms give in
Eqs.~(\ref{eq:RDconn0}) and (\ref{eq:RSconn0}). This is another
indication that the PQ theory is unphysical. The sum of the ratios is,
however, physical and should be an exponential. 
Indeed, we find that the quadratic term in \(R_\CD(t) +
R_\CS(t)\) is exactly that needed to give the quadratic term in the
expansion of Eq.~(\ref{eq:RD+RSlinear}). This provides a non-trivial
check on our results.

\bigskip
We now return to the terms linear in \(t\). We can first dispense with
the time-orderings of Figs.~\ref{fig:fig3}(c) and (d). Here a linear
term arises only when the pions in the loop have \(\vec q=0\). This
means that the contribution to \(C_{\CD,\CS}(t)\) is of order \(L^0\)
and thus that to \(R_{\CD,\CS}(t)\) is order \(1/L^6\), which is
higher order than we are controlling. The same suppression holds for
the contribution of pions with \(\vec q=0\) in time ordering
of Fig.~\ref{fig:fig3}(b). Thus we are left to consider the 
\(\vec q\ne 0\) contributions to the time-orderings of
Figs.~\ref{fig:fig3}(a) and (b).

Using Eq.~(\ref{eq:CAs}), the time ordering of Fig.~\ref{fig:fig3}(a)
gives the following contribution to \(R_\CD(t)\):
\begin{eqnarray}
R_\CD(t) &\supset& \tilde w_8'^2 \frac1{L^6} \int_0^t dt_2
\int_0^{t_2} dt_1 \int_{\vec x_1} \int_{\vec x_2} \nonumber\\ &&\quad
\frac{e^{-2M_\pi (t_1-t_2)}}{(2M_\pi)^2} \overline G_\pi(x_1,x_2)^2
\,,
\end{eqnarray}
where the bar over the pion propagator indicates that the \(\vec q=0\)
mode has been removed. We are interested in the coefficient of
\(t/(4M_\pi^2 L^3)\) for large \(t\), and so we again take 
a derivative with respect to \(t\) and send \(t\to\infty\). 
The result, after some manipulations, is \(\tilde w_8'^2 I_a\), with
\begin{equation}
I_a = \int_{-\infty}^{0} dt_1 \int_{\vec x_1} e^{-2 M_\pi t_1}\,
\overline G_\pi(x_1,0)^2 \,.
\end{equation}
Note that the growth of the exponential as \(t_1\) becomes more
negative is overwhelmed by the decrease in \(\overline G_\pi^2\).

A similar calculation for the time-ordering of Fig.~\ref{fig:fig3}(b)
yields, for the coefficient of \(t/(4 M_\pi^2 L^3)\), the result
\(\tilde w_8'^2 I_b\) with
\begin{equation}
I_b = \int_0^{\infty} dt_1 \int_{\vec x_1} e^{-2 M_\pi t_1}\, 
\overline G_\pi(x_1,0)^2 \,.
\end{equation}
Here both the exponential and the \(\overline G_\pi^2\) factors
decrease as \(t_1\) becomes larger. Combining the two time-orderings
we find \(\tilde w_8'^2\) times
\begin{align}
I_a+I_b & = \int d^4x_1\; e^{-2 M_\pi t_1} \bar G_\pi(x_1,0)^2 
\label{eq:I2a+b}
\\ 
&= I_2(s=4M_\pi^2)^{FV} \,.
\label{eq:I2sFV}
\end{align}
As noted in the second line, this integral is the finite volume
version of the corresponding infinite volume integral at threshold,
expressed in position space.  
The factor of \(\exp(-2 M_\pi t_1)\) simply leads to the 
injection of the physical
threshold four-momentum through the loop. 

If we could ignore the difference
\begin{equation}
\delta I_2 \equiv I_2(s=4M_\pi^2)^{FV}-I_2(s=4M_\pi^2) \,,
\end{equation}
the result (\ref{eq:I2sFV}) tells us that, in the coefficient of \(t/(4
M_\pi^2 L^3)\), all the NNLO corrections to \(\CD\) and \(\CS\)
proportional to \(I_2(s)\) appear exactly as in infinite volume. As we
will see shortly, however, \(\delta I_2 \propto 1/L\) (plus
exponentially suppressed terms) so we do need to calculate it.

First, however, we extend the calculation of the coefficient of \(t\)
to \(s\)-channel loops with momentum-dependent vertices---the loops which
give rise to the integrals \(I_4(s)\) and \(I_6(s)\). The steps
outlined above lead to the same combined integral (\ref{eq:I2a+b}),
except that there are two or four derivatives acting on the various
factors in the integrand. As was the case with the \(u\)-channel
diagrams, one must carefully handle the discontinuity at \(t_1=t_2\)
when two derivatives act on an internal propagator. The end result,
however, is still as claimed.  This allows us to rewrite all
contributions to the time dependent correlator which are linear in
\(t\) as the corresponding contributions to the finite volume
threshold amplitudes. Next we may relate these to the
\(I_2(s=4M_\pi^2)^{FV}\) and \(I_1^{FV}\) using the same expressions,
(\ref{eq:i4def}) and (\ref{eq:i6def}), as hold in infinite volume
(with \(s=4M_\pi^2\)). From this follows that the difference between
the finite and infinite volume forms is exponentially suppressed.
Thus, aside from the \(\delta I_2\) terms, we find that the
coefficient of \(t/(4 M_\pi^2 L^3)\) generated by \(s\)-channel loop
diagrams is just full infinite-volume contribution from the same
diagrams evaluated at threshold.

Our final task is to evaluate \(\delta I_2\). Standard manipulations
lead to the following expression
\begin{equation}
\delta I_2 = \left[ \frac1{L^3}\sum_{\vec q\ne 0} - \int
  \frac{d^3q}{(2\pi)^3} \right] \frac1{4 E_{\vec q}\, [\vec q \,]^2} \,,
\end{equation}
where \(E_{\vec q} = \sqrt{[\vec q \,]^2 + M_\pi^2}\). Since the UV divergences cancel in the difference between sum and
integral, we can introduce a regulator term \(\exp(-\alpha [\vec q \,]^2)\), 
as long as we send \(\alpha\to 0^+\). With this regulator in
place, we can use L\"uscher's summation formula~\cite{Luscher:1986pf}
(in the particular form quoted in Ref.~\cite{BGpipi})
\begin{equation}
\left[\frac1{L^3} \sum_{\vec q\ne0} - \int
  \frac{d^3q}{(2\pi)^3}\right] \frac{f(\vec q^2)}{\vec q^2} =
\frac{c_1 f(0)}{4 \pi L} -\frac{f'(0)}{L^3} \,,
\end{equation}
where \(c_1\) is the constant given after Eq.~(\ref{eq:Luscher}). This
result is valid up to exponentially small corrections as long as \(f\)
and all its partial derivatives are square integrable (as is the case
with our regularized sum). Applying this result, we find
\begin{equation}
\delta I_2 = \frac{c_1}{16 \pi M_\pi L} + \frac{1}{8 (M_\pi L)^3} \,.
\end{equation}
At the order we are working, we need keep only the \(1/L\) term.

Collecting all the \(1/L\) terms, we find their contribution to the
ratios to be:
\begin{align}
R_\CD(t) & \supset \left[\left( \tilde w_8' -
  \frac{4M_\pi^2}{3f^2}\right)^2 \!\!\!  + 4 w_6'^2\right]\!  \frac1{4
  M_\pi^2 L^3} \frac{c_1 t}{16 \pi M_\pi L} \\ R_\CS(t) & \supset 4
w_6' \left(\tilde w_8' - \frac{4 M_\pi^2}{3 f^2}\right) \frac1{4
  M_\pi^2 L^3} \frac{c_1 t}{16 \pi M_\pi L} \,.
\end{align}
The combinations of LECs here are the same as in the \(t^2\) terms,
Eqs.~(\ref{eq:RDquad}) and (\ref{eq:RSquad}), because the both arise
from the \(I_2(s)\) integral in finite volume.

A check on this result is that the \(t/L^4\) term in the
\(\pi^+\pi^+\) correlator \(R_\CD(t)+R_\CS(t)\) agrees with that
obtained with L\"uscher's general formula (\ref{eq:Luscher}).

\subsection{Summary of results}
\label{subsec:summary}

Collecting the results from this and the previous section, we find
\begin{equation}
\begin{split}
R_\CD(t) & = 1 + \CO\left(\frac{M_\pi^2}{f^2}\frac{1}{M_\pi^3 L^3}\right) + \frac{t}{4 M_\pi^2 L^3} \CD(4M_\pi^2,0,0) \\
& + \left[\left(\tilde w_8' - \frac{4 M_\pi^2}{3 f^2}\right)^2 + 4 w_6'^2\right] \\
& \hspace{20pt} \times \left\{\frac1{4 M_\pi^2 L^3}\frac{c_1 t}{16 \pi M_\pi L} + \left(\frac1{4 M_\pi^2 L^3} \right)^2 \frac{t^2}{2} \right\} \\
& \hspace{90pt} \times \Bigg \{ 1 + \CO\left(\frac{M_\pi^2}{f^2}\frac{1}{M_\pi L}\right) \!\! \Bigg \} \\
& \hspace{-30pt}+ {\cal O}\left(\left[\frac{M_\pi^2}{f^2} \frac{t}{M_\pi^2 L^3}\right]^3\right) + {\cal O}(e^{-M_\pi L}) + \textrm{exp. suppr.} \,,
\label{eq:RDfinal}
\end{split}
\end{equation}
Here $\CD$ is the full NNLO amplitude.
The corresponding result for \(R_\CS(t)\) is
\begin{equation}
\begin{split}
R_\CS(t) & =\CO\left(\frac{M_\pi^2}{f^2}\frac{1}{M_\pi^3 L^3}\right) + \CS(4M_\pi^2,0,0) \frac{t}{4 M_\pi^2 L^3} \\
& + 4 w_6' \left(\tilde w_8' - \frac{4 M_\pi^2}{3 f^2}\right) \\
& \hspace{20pt} \times \left\{\frac1{4 M_\pi^2 L^3}\frac{c_1 t}{16 \pi M_\pi L} + \left(\frac1{4 M_\pi^2 L^3} \right)^2 \frac{t^2}{2} \right\} \\
& \hspace{90pt} \times \Bigg \{ 1 + \CO\left(\frac{M_\pi^2}{f^2}\frac{1}{M_\pi L}\right) \!\! \Bigg \} \\
& \hspace{-30pt}+ {\cal O}\left(\left[\frac{M_\pi^2}{f^2} \frac{t}{M_\pi^2 L^3}\right]^3\right) + {\cal O}(e^{-M_\pi L}) + \textrm{exp. suppr.} \,.
\label{eq:RSfinal}
\end{split}
\end{equation}
Note that the last three lines of Eqs.~(\ref{eq:RDfinal}) and (\ref{eq:RSfinal}) are identical.

Thus, if one can measure the coefficients of \(t/L^3\), one obtains
the corresponding infinite volume PQ threshold scattering amplitudes
{\em at NNLO}. In fact, it is plausible that this holds to all orders,
since the analysis above shows how picking out the \(t\) term
corresponds to LSZ reduction in infinite volume. 
Of course, it is non-trivial to pick out these coefficients, but our
results provide the coefficients of the leading competing
terms---those proportional to \(t/L^4\) and \(t^2/L^6\).

For completeness we give the PQ results for the threshold scattering
amplitudes. Using the results \(F(4M_\pi^2)=0\), \(F(0)=-2\) and the
notation \(Q=\log(M_\pi^2/\mu^2)/(16 \pi^2)\) these are
\begin{multline}
\CD(4M_\pi^2,0,0) = 2 w_6' + \CD^{(1,2)}_{an}(4 M_\pi^2,0,0)\\
+ \frac{1}{16 \pi^2} \bigg[-\frac{M_\pi^4}{2f^4} + \frac{M_\pi^2}{f^2}(2 w_6'-7 w_8') -14 w_6'^2 - 10
w_6' w_8' \\- (17/2) w_8'^2 \bigg] + Q \bigg [ -\frac{5 M_\pi^4}{2 f^4}
  +\frac{M_\pi^2}{f^2}(4 w_6' + 2 w_7' -5 w_8') \\ -22 w_6'^2 - 10 w_6'
  w_8' - (21/2) w_8'^2\bigg ]
\label{eq:CDthresh}
\end{multline}
and
\begin{multline}
\CS(4M_\pi^2,0,0) =  w_8' -
\frac{M_\pi^2}{f^2} + \CS^{(1,2)}_{an}(4 M_\pi^2,0,0) 
- \frac{\delta M_{an}^2}{f^2}
\\ 
+ \frac{1}{16 \pi^2} \bigg[ \frac{M_\pi^4}{f^4} +
\frac{2M_\pi^2}{f^2}(-2 w_6'+3 w_8') - 4 w_6' w_8' + 5 w_8'^2 \bigg] 
\\ 
+Q \bigg [ -\frac{M_\pi^4}{ f^4} + 
\frac{M_\pi^2}{f^2}(2 w_6' - 2 w_7' + 8 w_8') 
\\ 
- 12 w_6' w_8' +5 w_8'^2 \bigg ] \,.
\label{eq:CSthresh}
\end{multline}
We do not give explicit expressions
for \(\CD^{(1,2)}_{an}(4 M_\pi^2,0,0)\) and
\(\CS^{(1,2)}_{an}(4 M_\pi^2,0,0)\) 
as they can be readily determined from
Eqs.~(\ref{eq:dan}) and (\ref{eq:san}).

\section*{Acknowledgments}

This work is supported in part by the US DOE grant
no.~DE-FG02-96ER40956. 

\appendix

\section{Additional contributions to \(\mathcal{D}\) 
and \(\mathcal{S}\) \label{app:indcon}}

In this appendix we analyze the additional contributions to the PQ
amplitudes that come from the \(a^3\) terms in the NLO Lagrangian and
from the \(m a^2\), \(a^4\) and \(p^2 a^2\) terms in the NNLO. We
begin by enumerating all of the \(a^3\) terms allowed by chiral
symmetry:
\begin{equation}
\begin{split}
\mathcal L_{a^3} 
& \sim \hat a^3 \langle \Sigma + \Sigma^\dagger \rangle^3 
\\ & {} + \hat a^3 \langle \Sigma - \Sigma^\dagger \rangle^2
                   \langle \Sigma + \Sigma^\dagger \rangle 
\\ & {} + \hat a^3 \langle \Sigma^2 + (\Sigma^\dagger)^2 \rangle 
                   \langle \Sigma + \Sigma^\dagger
\rangle 
\\ & {} + \hat a^3 \langle \Sigma^2 - (\Sigma^\dagger)^2 \rangle 
                   \langle \Sigma - \Sigma^\dagger \rangle 
\\ & {} + \hat a^3 \langle \Sigma^3 + (\Sigma^\dagger)^3 \rangle 
\\ & {} + \hat a^3 \langle \Sigma + \Sigma^\dagger \rangle \,.
\end{split}
\end{equation}
We use \(\sim\) throughout this appendix to indicate that the
two sides are equal with an independent LEC multiplying each term.
The key observation is that these terms, 
when expanded in \(\mathcal \pi\), produce 
\(\langle \pi^2 \rangle\), \(\langle \pi^2 \rangle^2\) and
\(\langle \pi^4 \rangle\) with independent coefficients. The
particular forms of these coefficients in terms of the unknown LECs
provides no useful information.

Both the $m a^2$ and $a^4$ sectors generate the same three
pion terms with independent coefficients.
The argument for the $m a^2$ terms is identical to that
for the $a^3$ sector, since
the spurions \(A\) and \(\chi\) transform in the same way.
For the $a^4$ sector we can show this result by displaying
three chiral operators which give linearly independent contributions to
the three pionic operators.
An example is 
\begin{equation}
\begin{split}
\mathcal L_{a^4} & \sim \hat a^4 \langle \Sigma + \Sigma^\dagger
\rangle^4 \\ & {} + \hat a^4 \langle \Sigma - \Sigma ^\dagger
\rangle^2 \langle \Sigma + \Sigma ^\dagger \rangle ^2 \\ & {} + \hat
a^4 \langle \Sigma + \Sigma^\dagger \rangle ^2 + \cdots \,,
\end{split}
\end{equation}
where the dots indicate additional terms which we do not need
to enumerate.

It remains to consider the \(p^2 a^2\) terms. Starting on the level of
the pion fields, we first list all two-derivative quadratic and
quartic terms. This is done without regard to chiral symmetry, using
only that \(\langle \pi \rangle=0\).
The resulting set is
\begin{equation}
\label{eq:dpdp}
\begin{array}{c}
\langle \partial_\mu \pi \, \partial_\mu\pi \rangle 
\\[5pt] \langle \partial_\mu \pi \, \pi \rangle
        \langle \partial_\mu \pi \, \pi \rangle
\\[5pt] \langle \partial_\mu \pi\,\partial_\mu \pi \, \pi^2 \rangle 
\\[5pt] \langle \partial_\mu \pi \, \pi \, \partial_\mu\pi \, \pi \rangle 
\\[5pt] \langle \partial_\mu\, \pi\partial_\mu \pi \rangle 
         \langle \pi^2 \rangle \,.
\end{array}
\end{equation}
Because this is the maximal set of pionic
terms to the order we are working, it is sufficient to show that
the \(p^2 a^2\) terms allowed by chiral symmetry produce the entire
set, with independent coefficients. This is indeed the case, as
follows from
\begin{equation}
\begin{split}
\mathcal L_{p^2 a^2} & \sim \hat a^2 \left [ \langle \partial_\mu \Sigma
\rangle^2 + \langle \partial_\mu \Sigma^\dagger \rangle^2 \right ] 
\\ & {}
+ \hat a^2 \langle \partial_\mu \Sigma \partial_\mu \Sigma ^\dagger \rangle 
\\
& {} + \hat a^2 \langle \partial_\mu \Sigma \partial_\mu \Sigma^\dagger
\rangle \langle \Sigma + \Sigma ^ \dagger \rangle^2 
\\ & {} + \hat a^2
\langle \partial_\mu \Sigma \partial_\mu \Sigma ^\dagger ( \Sigma^2 +
(\Sigma^\dagger)^2 ) \rangle 
\\ & {} + \hat a^2 \langle \partial_\mu
\Sigma \partial_\mu \Sigma (\Sigma^\dagger)^2 \rangle + \cdots \,.
\end{split}
\end{equation}
These terms are enough to independently give the set (\ref{eq:dpdp}).

From these results, we can determine
the contribution of all \(a^3\), \(a^2 m\), \(a^4\) and
\(p^2 a^2\) terms to the mass and wave function renormalizations and to
the PQ amplitudes. We find
\begin{align}
\delta z_{ad} & \sim \hat a^2 \\ \delta M^2_{ad} & \sim \hat a^2 M_0^2
+ \hat a^3 + \hat a^4 \\
\label{eq:dad}
\mathcal D_{ad}(t) & \sim \hat a^2 t + \hat a^2 M_0^2 + \hat a^3 +
\hat a^4 - 2 \delta z_{ad} \mathcal D^{(0)} \\
\label{eq:sad}
\begin{split}
\mathcal S_{ad}(s) & \sim \hat a^2 s + \hat a^2 M_0^2 + \hat a^3 +
\hat a^4 - 2 \delta z_{ad} \mathcal S^{(0)}(s) \\ & {} + 2 \delta
M_{ad}^2/(3f^2) \,.
\end{split}
\end{align}

\section{Determining $W'_7$}
\label{app:w7}

In this appendix we sketch a method for determining the LEC $W'_7$
in which its contribution appears at tree-level.
The method is by no means unique, but it is the simplest
approach we have found within the context of pion scattering.

Expanding out the $W'_7$ term in the LO chiral Lagrangian (\ref{eq:lolag}),
the first non-vanishing term is 
\begin{equation}
\CL_{\rm LO} \supset \frac{w'_7}{18 f^2} \langle \pi^3\rangle^2
\,.
\label{eq:w7vertex}
\end{equation}
To get a tree-level contribution from this vertex one needs
six external pions.
We propose calculating the following finite-volume correlation function
\begin{eqnarray}
C_{3\pi}(t) = \langle 
\tilde \pi_{12}(0) \tilde \pi_{23}(0) \tilde \pi_{45}(0)
\tilde \pi_{31}(t) \tilde \pi_{56}(t) \tilde \pi_{64}(t)
\rangle
\,,
\label{eq:C3pidef}
\end{eqnarray}
and then forming the ratio
\begin{equation}
R_{3\pi}(t) = \frac{C_{3\pi}(t)}{C_\pi(t)^3}\,.
\end{equation}
Here $\tilde\pi_{jk}$ are the $\vec p=0$ fields
defined in Eq.~(\ref{eq:pitildedef}) and the single-pion
correlator $C_\pi(t)$ is defined in Eq.~(\ref{eq:Cpidef}).
The subscripts on the pion fields indicate valence flavors,
of which there must be six.
We consider only $t>0$ in the following.

\begin{figure}[btp!]
{\includegraphics[angle=270,width=0.4\textwidth]{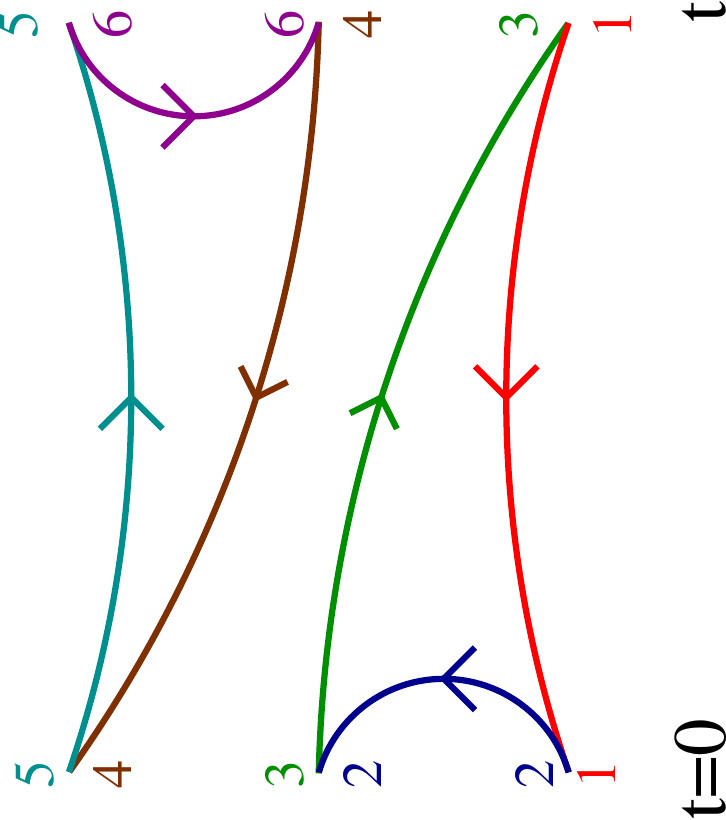}}
\caption{
Quark contraction for the correlator $C_{3\pi}$
of Eq.~(\ref{eq:C3pidef}).
\label{fig:w7fig1}}
\end{figure}

The choice of valence fields in (\ref{eq:C3pidef}) allows
only a single quark-level contraction, shown in Fig.~\ref{fig:w7fig1}.
By construction, this contraction has two quark loops,
in order to match with the
double-strace six-pion vertex of (\ref{eq:w7vertex}).
This will be a more challenging contraction to calculate in numerical
simulations than
those of Fig.~\ref{fig:contr}, because of the ``source to source''
propagators at times $0$ and $t$. Nevertheless, with recent advances
in calculations of ``all-to-all'' propagators, we expect that
the calculation should be feasible.

We have written the correlator in terms of the pion fields from
the chiral Lagrangian, but, as discussed in Sec.~\ref{subsubsec:tadpole},
in practice one would use a quark-level pseudoscalar field,
$\bar q_k \gamma_5 q_j$. The corresponding chiral operator is
\begin{equation}
(\Sigma-\Sigma^\dagger)_{jk}= c \left[ \pi_{jk}
-\frac{1}{3 f^2} (\pi^3)_{jk}+ \dots \right]\,,
\label{eq:interp}
\end{equation}
where the constant $c$ is known but not needed.
It turns out that in this calculation, unlike that in the main text,
the $\pi^3$ part of the interpolating operator
contributes at leading order to the quantities of interest.
This means that it is essential for the present method to use
(a discretization of) a local pseudoscalar bilinear to create
the pion fields, and not, for example, a non-local operator.

\begin{figure}[tbp!]
{\includegraphics[angle=270,width=0.45\textwidth]{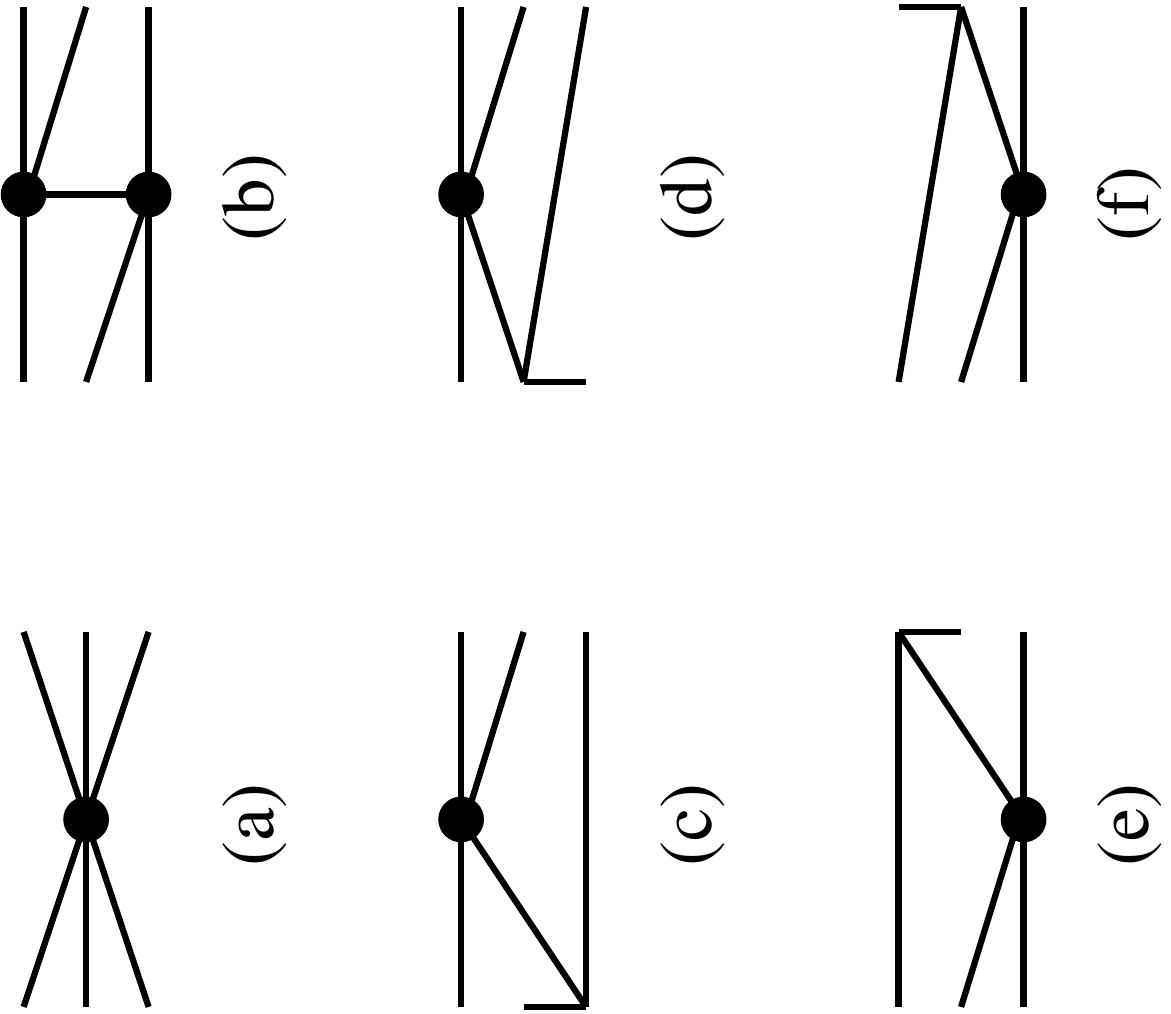}}
\caption{
Tree-level diagrams in PQWChPT contributing to linear or quadratic
dependence in $C_{3\pi}(t)$. The flavor indices of the external fields
are (implicitly) ordered as in Fig.~\ref{fig:w7fig1}.
Notation for vertices as in Fig.~\ref{fig:infvol}.
\label{fig:w7fig2}}
\end{figure}

Our choice of flavor indices significantly restricts the diagrams that
can appear and the vertices that contribute.
There is no pion-disconnected diagram, and the
two tree-level diagrams which contribute are shown
in Figs.~\ref{fig:w7fig2}(a) and (b). 
Note that the four-pion vertices in (b) must be attached to the
external legs as shown, other possibilities being forbidden by the
flavor indices.
There are also contributions involving the $ \pi^3$ 
and $ \pi^5$ parts of interpolating fields,
Eq.~(\ref{eq:interp}). For reasons explained below, the only
diagrams of this type contributing to quantities of interest
are those shown in Figs.~\ref{fig:w7fig2}(c-f).
We stress that all six diagrams in Fig.~\ref{fig:w7fig2} contribute
at the same order in WChPT.

We begin by discussing the three-pion scattering diagram (a).
Only the $w'_7$ vertex in $\CL_{\rm LO}$ has the
$\langle \pi^3\rangle^2$ form needed to contribute.
A straightforward calculation along the lines of those discussed
in the main text leads to
\begin{equation}
R_{3\pi}(t)\Big|_{w'_7} = -\frac{w'_7}{f^2} \frac1{L^6 (2M_0)^3} 
\left(t + \frac{1}{3M_0}\right)
\,.
\label{eq:Rw7}
\end{equation}
The term linear in $t$ arises, as usual, because the interaction
can occur at any time in the range $0-t$.
The volume suppression is now $1/L^6$ 
[compared to $1/L^3$ for the two-pion interaction, 
as in Eq.~(\ref{eq:RDLO})]
because all three pions must be in contact.
Note that at the order we are working in this appendix,
$M_0$ and $M_\pi$ are interchangeable.

Turning now to Fig.~\ref{fig:w7fig2}(b), we find that only
vertices having the pionic form $\langle \pi^4\rangle$ contribute. 
Such a form arises only from the mass, $w'_6$ and $w'_8$ terms
in $\CL_{\rm LO}$ [eq.~(\ref{eq:lolag})].
The kinetic term does not contribute.
After a straightforward but tedious calculation, we find\footnote{%
Recall that
$\widetilde w'_8= w'_8 + \frac{M_0^2}{3f^2}$
and 
$\frac{M_0^2}{f^2}=\frac{\chi}{f^2}+ 2 w'_6+w'_8$.}
\begin{eqnarray}
\lefteqn{R_{3\pi}(t)\Big|_{(w'_8)^2} = \frac{-9 (\widetilde w'_8)^2}{2}
\frac{1}{L^6 (2 M_0)^4}}
\nonumber\\
&&\times
\left(\frac{t^2}2 + \frac{t}{M_0}
+ \frac{11}{48 M_0^2}
+ \frac{e^{-2 M_0 t}}{16 M_0^2}\right)
\,.
\label{eq:Rw82}
\end{eqnarray}
In the LCE regime, $w'_k \sim M_0^2/f^2$, so the term
linear in $t$ is of the same order as that in the
$w'_7$ contribution (\ref{eq:Rw7}).

The quadratic term in (\ref{eq:Rw82}) arises because of the
presence of two vertices, much like the quadratic term
discussed in the main text.
The linear term arises both from contributions in which the
two vertices are close in time and integrated together from $0-t$,
and when one of the vertices is close to either $0$ or $t$.
The latter origin suggests that one should also consider
contributions in which one of the vertices is ``absorbed''
into the sources. Indeed, such diagrams, exemplified by
those in Fig.~\ref{fig:w7fig2}(c-f), do contribute at the same order.

As is evident from the result (\ref{eq:Rw82}) there are exponentially
falling terms for small $t$. There are also contributions from 
excited states which show up in ChPT as contact terms.
To avoid these, we consider henceforth only terms proportional
to $t$ and $t^2$. It turns out that there are no other LO diagrams
leading to quadratic dependence, and the only other diagrams leading
to linear dependence are those of Fig.~\ref{fig:w7fig2}(c-f).
Evaluating these diagrams, we find
\begin{equation}
R_{3\pi}(t)\Big|_{w'_8} = 6 \frac{\widetilde w'_8}{f^2}
\frac{1}{L^6 (2 M_0)^3}  t + {\rm const.}
\,.
\label{eq:Rw8}
\end{equation}

Combining these results gives
\begin{eqnarray}
\lefteqn{R_{3\pi}(t) \supset 
-\frac{t^2}{(2M_0)^4 L^6} \frac{9 (\widetilde w'_8)^2}{4}}
\nonumber\\
&& - \frac{t}{(2M_0)^3 f^2 L^6} \left[w'_7-6 \widetilde w'_8
+\frac94\frac{f^2}{M_0^2} (\widetilde w'_8)^2\right]\,.
\label{eq:R3pi}
\end{eqnarray}
Higher order contributions will lead to corrections suppressed
powers of $M_\pi/(4\pi f)$, $a$ and $1/L$.
Assuming these corrections to be small, the coefficient
of $t^2$ allows one to determine $\widetilde w'_8$, 
while that of $t$ gives a combination of $w'_7$ and $\widetilde w'_8$.
Alternatively, $\widetilde w'_8$ can be determined from other quantities
such as the two pion correlators discussed in the main text.
Either way, given $\widetilde w'_8$ one can use $R_{3\pi}(t)$ to
determine $w'_7$.

A noteworthy feature of the result (\ref{eq:R3pi}) is that
the $t^2$ term becomes comparable to the
linear term at the relatively short time $t\sim 1/M_0$.
This differs from the
corresponding results for the two-pion correlators 
[Eqs.~(\ref{eq:RDfinal}) and (\ref{eq:RSfinal})]
for which the quadratic term becomes important at 
\begin{equation}
t \sim \frac1{M_0} (M_0 L) (f L)^2 \gg \frac1{M_0}
\label{eq:longt}
\end{equation}
[see Eq.~(\ref{eq:noquadcond})].
The last inequality follows since one must have
$M_0 L\gg 1$ and $f L \gtrsim 1$ to avoid large finite-volume effects.
The upshot of this discussion is that the quadratic term
is much more important for the three-pion ratio than for the
two-pion ratios.
This is not true for cubic and higher order terms,
which become important only at the longer times of
Eq.~(\ref{eq:longt}).
%

\bibliographystyle{apsrev} 
\bibliography{ref} 

\end{document}